# A method for constructing an interplanetary trajectory of a spacecraft to Venus using resonant orbits to ensure landing in the desired region

**Vladislav A. Zubko*, (1), Natan A. Eismont (1), Konstantin S. Fedyaev (1,2), Andrey A. Belyaev (1)**

(1) *Space Research Institute of the Russian Academy of Sciences, 84/32 Profsoyuznaya Str., Moscow, 117997, Russian Federation*
(2) *Moscow Aviation Institute (National Research University), 4 Volokolamskoe shosse, Moscow, 125993, Russian Federation*

**Abstract:** A problem of constructing the trajectory of a spacecraft flight to Venus within the framework of a mission including landing of a lander in a given region of the planet's surface is being considered. A new celestial mechanics related method based on the use of gravity assist maneuver near Venus is proposed to transfer the spacecraft to a heliocentric orbit resonant with the orbit of Venus, so that, at the next approach the planet, the given region of the surface becomes attainable for landing. It is shown that the best resonant orbit in terms of the cost of the characteristic velocity is an orbit with a 1:1 ratio of the period to the orbital period of Venus. A procedure for choosing one of possible resonant orbits depending on coordinates of the desired landing point on the surface and the launch date of the mission is described. An example of calculating the flight trajectory that ensures landing in the Vellamo-South region of the Venus surface at launch from the Earth in 2031 is considered.

## Nomenclature

$\Delta V_0$ = characteristic velocity required for the launch from the low-Earth orbit to Venus.
$\Delta V_1$ = characteristic velocity required for the flight on Venus-Venus arc.
$\Delta V_\Sigma$ = total characteristic velocity required for the flight to Venus.
$V_{res}$ = velocity of the spacecraft on a *m:n* resonant circular heliocentric orbit.
$\mathbf{V}_r$ = vector of the spacecraft asymptotic velocity at Venus in the planetocentric ecliptic reference frame.
$V_r$ = magnitude of the vector $\mathbf{V}_r$.
$\mathbf{V}_a$ = vector of the heliocentric velocity of the spacecraft at approaching Venus in the heliocentric ecliptic reference frame.
$\mathbf{V}_p$ = vector of the Venus orbital velocity in the heliocentric ecliptic reference frame.
$\mathbf{V}_r^-$ = vector of the incoming asymptotic velocity of the spacecraft at Venus in the planetocentric ecliptic reference frame.
$\mathbf{V}_r^+$ = vector of the outgoing asymptotic velocity of the spacecraft at Venus in the planetocentric ecliptic reference frame.
$\alpha$ = angle between $\mathbf{V}_r^-$ and $\mathbf{V}_r^+$;. $\alpha_{min}$, $\alpha_{max}$ = minimum and maximum turn angles required to transfer to a given *m:n* resonant orbit.
$\alpha^*$ = natural turn angle.
$\delta$ = angle between $\mathbf{V}_p$ and $\mathbf{V}_r^-$.
$\gamma$ = angle between pole of the Venus and projection of the $\mathbf{V}_r^+$ on the plane perpendicular to $\mathbf{V}_p$.
$\Phi$ = angle between $\mathbf{V}_r^+$ and $\mathbf{V}_p$.
$r_\pi$ = periapsis distance of the spacecraft trajectory at Venus flyby.
$\mathbf{r}_p$ = radius vector of Venus in the heliocentric ecliptic reference frame.
$\psi$ = angular radius of landing circle.

$\varpi$ = angular radius of the circle of possible periapsis's.
$\nu$ = module of the true anomaly of the spacecraft at the entry point.
$\theta$ = horizontal flight path angle (entry angle).
$t_o, t_1, t_k$ = moments of launch of the spacecraft from a low-Earth orbit, the 1st flyby of Venus, the landing, respectively;
$L$ = landing point on the surface of Venus
$\lambda_L$, $\varphi_L$ = longitude and latitude of the required landing point *L;*
$\lambda$ , $\varphi$ = planetocentric longitude and latitude;
$P(\varphi_P, \lambda_P)$ = point of virtual intersection of the Venusian orbital velocity vector $\mathbf{V}_P$ with the sphere that represents the planet's surface;
σ = angular distance between $L(\varphi_L, \lambda_L)$ and $P(\varphi_P, \lambda_P)$.
$\mu = 324859\ km^3/s^2$ – gravitational parameter of Venus.
$\mu_{sun} = 132\,712\,440\,018\ km^3/s^2$ – gravitational parameter of the Sun.
***Notice:*** *if the upper script "LO" is used then the vector is given in the local orbital coordinate system (description given in section 1.4).*

# Introduction

Venus has always been a significant object for space exploration. Almost all the missions launched to Venus were a huge success. For example, in the USSR, 13 of total 18 successful Venusian missions (16 *Venera* missions and 2 *Vega* missions) ended up with a descent in the atmosphere of Venus, and 10 vehicles landed on its surface (Glaze et al. 2018; Colin 1977; Taylor 2014). Mariner-2 (1962, USA/NASA) was the first spacecraft that successfully reached Venus in December 1962 at an altitude of about 34 thousand km and transmitted data on the hot atmosphere of Venus, as well as on the absence of its own magnetosphere (Neugebauer and Snyder 1966). It is also worth highlighting the flight of the Mariner-10 spacecraft (1973, USA/NASA) aimed to study Mercury from a flyby trajectory (Shirley, 2003). During the flight to Mercury, Mariner-10 approached Venus at an altitude about 5770 km in order to use its gravity field to lower orbital energy and transfer to a flyby trajectory to Mercury. It is also necessary to highlight two NASA missions of the Pioneer-Venus series (1978, USA/NASA), in which a probe was dropped on Venus, although it did not reach the surface. Note that the last two missions launched in the USSR (*Vega-1* and *Vega-2*, 1984), after exploring Venus and dropping the probe to its surface, were targeted at Halley's comet (1P/Halley)[1] using a gravity assist maneuver near Venus. From 1989 to 1994 a spacecraft of the Magellan mission (1989, USA/NASA) conducted a full-scale radar mapping of the Venus surface. Among Venusian missions carried out after 2000 for studying the surface and atmosphere of Venus from an orbiter there were the Venus Express mission (2005, EU/ESA)[2] (Glaze et al. 2018), as well as *PLANET-C* also known as Akatsuki (2010, Japan/JAXA) [3].

An interest to explore Venus grew up significantly after 2020 when a team of scientists found traces of phosphine at heights of 50-60 km above the surface (Greaves, Richards, Bains, Rimmer, Sagawa, et al. 2021) when analyzing an atmosphere spectrum on observations by the Atacama Telescope (Chile) and the James Clark Telescope (Hawaii). The detection of phosphine traces may indirectly indicate the presence of life in the Venusian atmosphere (Limaye et al. 2018), which has been discussed for many years. It is worth noting that the results of the work (Greaves, Richards,

* **Corresponding author**: Vladislav Zubko, Ph.D. student, researcher at the Space Research Institute of the RAS, 117997, Moscow, Profsoyuznaya Street 84/32, tel: +7(916)-922-4158; e-mail: v.zubko@iki.rssi.ru
[1] The *Vega-1* and *Vega-2* spacecraft were able to get closer to the nucleus of the comet and, despite of a large miss from the center of the nucleus due to the inability to determine the exact nucleus position in the coma, the trajectory measurements made by the soviet *Vega* allowed to clarify the nucleus coordinates, so that, based on the obtained parameters, the trajectory of the European *Giotto* spacecraft was adjusted to ensure a close approach the comet nucleus at a distance of about 200-700 km.
[2] The section on the website of the European Space Agency, ESA, dedicated to the project: https://blogs.esa.int/rocketscience/2015/01/23/venus-express-the-last-shout/ (date of application: August 20, 2022).
[3] Official web page of the project: https://akatsuki.isas.jaxa.jp/en/mission/ (date of application: August 20, 2022).

Bains, Rimmer, Sagawa, et al. 2021), (Greaves, Richards, Bains, Rimmer, Clements, et al. 2021) are widely discussed in the scientific community, and as of 2022 there are also other theories offering alternative explanations of the obtained results. For example, in the paper (Villanueva et al. 2021) it is claimed that traces of phosphine were in fact sulfuric acid.

The possibility of life's existence in the Venusian clouds, despite all the controversy of such a hypothesis, has spurred the interest of the world's leading space countries in conducting further studies of Venus. Currently, six scientific missions to the planet are under development. Two of them, *DAVINCI+* (NASA) and *Venera-D* (Roscosmos) suppose sending a lander to the surface of Venus, while others involve conducting studies of Venus by sending spacecraft to a near-Venus orbit.

Selection of a landing site for a Venus exploration mission is crucial for its success from both an engineering and a scientific points of view (Ivanov et al. 2021; Ivanov, Zasova, Gerasimov, et al. 2017). Mainly this selection is determined by the criteria of landing safety and scientific significance, but it is also limited by the required ballistic parameters of the flight trajectory. And here there is a contradiction between the scientific significance of the particular landing areas and ensuring safety of the landing vehicle. Thus, according to (Ivanov et al. 2021; Ivanov, Zasova, Gerasimov, et al. 2017), from a scientific point of view, tesserae, ancient formations that are the only "window" into the past of Venus, have the highest priority for landing among all the surface areas. However, landing there is extremely difficult because the surface of the tesserae is dissected by many low and high ledges. Their lower parts are probably covered with scree while the upper ones are rock walls close to vertical. Plain places of the tesserae do not exceed several kilometers in diameter (Ivanov et al. 2021; Ivanov, Zasova, Gerasimov, et al. 2017). Clusters of volcanic rocks and areas of active volcanoes may also be considered as sites of a high scientific significance since their study is critically important for understanding the geology of Venus (D'Incecco et al. 2021), but they also seem to be difficult to perform landing in. On the other hand, lowland areas, ridge plains or ridge belts, which are safe for landing, do not have a high scientific significance (Ivanov et al. 2021; Ivanov, Zasova, Gerasimov, et al. 2017).

Difficulty of selecting landing sites appears, among other reasons, due to peculiarities of motion of Venus. A period of the own rotation of the planet is equal to 243 Earth days, and a spacecraft transfer from the Earth to Venus is limited by the costs of the characteristic velocity, its launch is possible, as a rule, during an interval of ±1 week up to ±3 weeks from the optimal launch date. As a result, the part of the planet's surface attainable for landing is limited to only a small area, which is less than 2% of the entire surface. If the mission consists of an orbiter operating simultaneously with a lander there are additional ballistic constraints appear. Those constraints lead to a greater reduction in those areas on the surface of Venus where landing can be carried out. The above-mentioned circumstances indicate the importance of detailed consideration of the issue of constructing a spacecraft trajectory that, on the one hand, would satisfy all ballistic constraints, and, on the other hand, would ensure landing in the area of the planet's surface that is the most preferable from a scientific point of view.

In this paper, we propose a method for constructing a spacecraft transfer trajectory to Venus, which provides landing of a lander in the desired region located in almost any area of the planet's surface. The method is based on the use of a Venus gravity assist and resonant orbits[4]. The gravity assist maneuver is performed by the spacecraft at the first approach Venus and serves for its transition to the required resonant orbit. The spacecraft landing in the desired region of the planet's surface is ensured by selecting a resonant orbit that allows landing in this region after the next approach Venus, i.e. after *m* revolutions of the spacecraft and *n* Venusian years.

[4] A spacecraft orbit resonant with the orbit of the planet in the ratio *m*:*n* (hereinafter, for brevity, we call such an orbit a *m*:*n* resonant one) is understood in this work as a heliocentric orbit of the spacecraft, the ratio of the period of which to the period of the orbit of the planet is a rational number *m*:*n*.

The use of resonant orbits in the proposed method is based on existing methods related to the use of resonant orbits in other problems of celestial mechanics. Methods using resonant orbits are of practical importance, for example, in the mission design to moons of Saturn or Jupiter. It is worth mentioning the work (Strange, Russell, and Buffington 2008), which proposes the use of the V-infinity globe method to obtain resonant orbits due to gravity assist maneuvers, and also demonstrates the application of the method for flights in the Saturn moons system. There are known works (Uphoff, Roberts, and Friedman 1976; Golubev et al. 2020; Golubev, Grushevskii, Kiseleva, et al. 2019) in which the authors apply gravity assist maneuvers at the main moons of Jupiter to reduce the orbital energy of the spacecraft relative to Jupiter. It is also worth mentioning techniques to obtain orbits of high inclination to the ecliptic, which are based on the use of gravity assist maneuvers near Venus (Golubev, Grushevskii, Koryanov, et al. 2019a; Golubev et al. 2018; Golubev, Grushevskii, Koryanov, et al. 2019b; Golubev et al. 2017). One more significant example of using resonant orbits is the Europa Clipper project. In this project gravity assist maneuvers at Europa were used to choose a resonant orbit for the mission spacecraft in order to provide the full coverage of Europa surface for its mapping (Campagnola et al. 2019).

Note that there are also other ways to increase efficiency of extending capabilities of landing on the planet's surface. The most effective ones include changing the aerodynamic shape of a lander, thus increasing its maneuverability during the descent. Such studies for missions to Venus as well as its effectiveness have been done in (Kosenkova 2021; Kosenkova and Minenko 2020; Vorontsov et al. 2011).

Earlier the authors conducted research on the development of this method (Eismont, Koryanov, *et al.*, 2021;. Eismont *et al.*, 2021; Eismont *et al.*, 2022). In the first two of the mentioned works, the use of resonant orbits to expand possibilities of reaching landing areas within the framework of the Russian Venera-D project was considered. In the research (Eismont, Nazirov, *et al.*, 2021; Eismont *et al.*, 2022), an analysis was made of the influence of the angle of entry into the atmosphere on the attainable landing areas within the scenario of a flight to Venus with a gravity assist maneuver.

In this paper, the previously considered method of expansion landing areas is generalized to the method for constructing flight trajectories to Venus, providing landing in a given area on its surface. The algorithm of transition the spacecraft to the required resonant orbit providing landing at a given point on the surface of Venus is considered. As a practical example, we describe a design of a flight trajectory to Vellamo-South (29°S, 164°E) which is one of the most interesting regions to the science.

# 1. Methods

## 1.1. Flight trajectory design

A standard approach for constructing calculated trajectories of interplanetary flights is to divide the spacecraft motion into separate sections so that in each section this motion can be approximately described by the keplerian model. During the passage of the spacecraft in the spheres of influence (SOI) of the planets, it is assumed that only the gravity force of the corresponding planet affects the trajectory of the spacecraft, and outside the SOI (on the heliocentric sections of the trajectory) only the gravity force of the Sun is taken into account, all other factors affecting the spacecraft (such as gravity attraction of other massive celestial bodies, polar compression of planets, pressure of the sunlight, etc.) are neglected. In addition, when the spacecraft moves in the SOI of a planet, its size is considered infinite (i.e. the planetocentric velocity of the spacecraft at the boundary of the SOI is assumed to be equal to its asymptotic velocity), and on heliocentric sections of the trajectory SOI of planets are considered as point-like (which is permissible in the first approximation due to the smallness of sizes of these spheres compared to heliocentric distances); accordingly, the time of flight through the SOI is neglected[5].

An approximate calculation of the spacecraft trajectory to Venus including a gravity assist maneuver and a flight along a resonant orbit can also be performed using the same method, but in this case, two heliocentric sections should be considered: flight from the Earth to Venus and then a subsequent flight from Venus to Venus in a resonant orbit. Thus, the trajectory calculation includes the following stages, illustrated in Fig. 1 (Eismont et al. 2022b; Eismont et al.. 2021c; Eismont, Nazirov, et al. 2021a):

(I) launch a spacecraft from a low-Earth orbit and flight into a SOI of the Earth;
(II) heliocentric flight from the Earth to Venus;
(III) flight in the SOI of Venus and a gravity assist maneuver at Venus, which transits the spacecraft into a heliocentric orbit resonant with the orbit of Venus;
(IV) heliocentric flight in a resonant orbit until the 2nd approach Venus;
(V) 2nd approach Venus with subsequent landing on its surface.

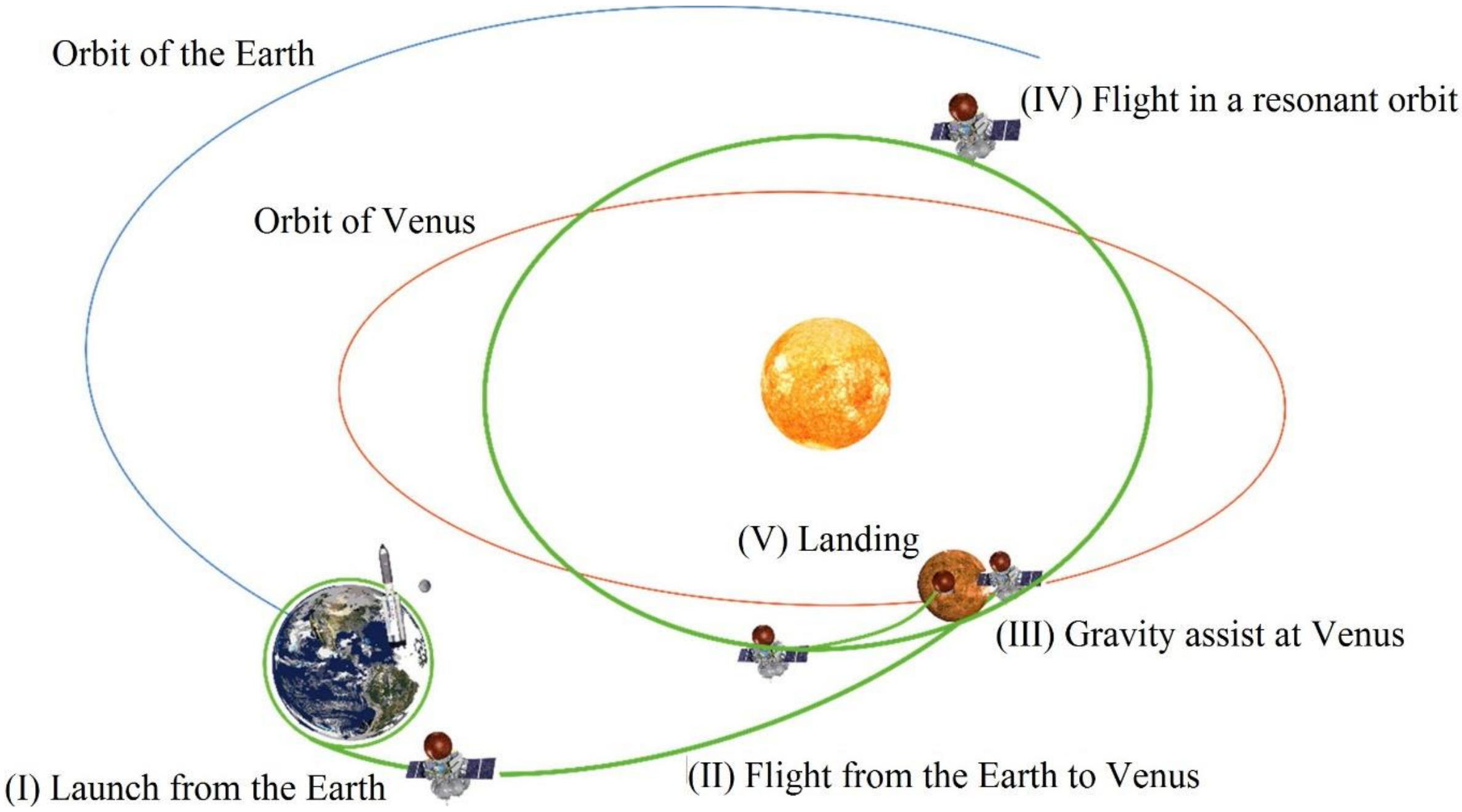

**Fig.1.** Stages of the spacecraft flight from the Earth to Venus in the mission scenario using a gravity assist maneuver and a flight in a resonant orbit. Notice that the example scenario is given for the 1:1 resonant orbit, but it also can be applied to any chosen *m:n* resonance.

Notice that the scheme presented in Fig. 1 is approximate because the stages (III)-(IV) may be repeated as many times as necessary to reach the required point on the Venus surface. However, every repetition of the stages (III)-(IV) would eventually lead to increase the total time of flight on $n$ periods of Venus.

The first step made at stages (I)-(II) is the construction of the trajectory of the flight from the Earth to Venus by solution of the Lambert problem. Note that in this paper the Lambert problem was solved using the Izzo method (Izzo 2015).

To calculate a trajectory of the spacecraft between each pair of celestial bodies, the patched conic method can be used. Thus, within the framework of the patched conics, the calculation of the spacecraft trajectory in the gravitational field of $N$ celestial bodies is reduced to $N$ problems of two

[5] In the complicated version of the patched conic approximation sizes of the SOI of the planets are assumed to be finite (Li, Zhao, and Li 2018)

bodies, which solutions are keplerian orbits (i.e., conic sections). The problem is to find these Keplerian orbits and to connect them (to patch) into the common resulting trajectory. Determination of the heliocentric sections of the spacecraft flight between each pair of celestial bodies is carried out by solving the Lambert problem, which consists of finding the trajectory of the spacecraft by the initial and final positions and the flight duration. Also, it should be emphasized that each part of the patched trajectory of the spacecraft is a result of ballistic (non-propelled) motion.

It is worth noting that the method of patched conic approximation significantly simplifies the analysis and optimization of flight trajectories and at the same time provides acceptable accuracy at the initial stages of mission development (Prado 2007; Uphoff, Roberts, and Friedman 1976; Li, Zhao, and Li 2018; Breakwell and Perko 1966).

The resulting trajectory can be refined by numerical integration into a more complete model of spacecraft motion in the central gravitational field of the Sun in presence of gravitational perturbations from other *N*–1 celestial bodies of the Solar System (Vallado, 2016).

The curve formed by the intersection of the bunch of incoming trajectories with the surface of the planet (in our case with planet's atmosphere) is a circle centered at the intersection point of the vector of incoming asymptotic velocity of the spacecraft and the surface of the planet. The points lying on this circle are considered as possible landing points. Then the radius of such a circle can be calculated as follows (Vallado 2016; Eismont et al. 2022):

$$\psi = \nu_{Entry} + \varpi, \tag{1}$$

where $\varpi$ is defined by formula (Vallado 2016; Eismont et al. 2022):

$$\varpi = \arccos\left(\frac{\mu}{\mu + r_\pi V_r^2}\right). \tag{2}$$

Notice that a part of the trajectory from the point of the entry to the Venusian atmosphere to the landing point should be calculated by numerical integration.

### 1.2. Transition of the spacecraft to a resonant orbit using gravity assist of Venus

In this subsection we consider in more detail a gravity assist maneuver at the first approach Venus and the subsequent transition of the spacecraft into a resonant orbit, which makes up the stage (III) in the scheme described above.

Let us determine incoming ($\mathbf{V}_r^-$) and outgoing ($\mathbf{V}_r^+$) asymptotic velocities of the spacecraft at the boundary of the Venus SOI from the velocity triangle (see Fig. 2), taking into account assumptions concerning the used model:

$$\begin{cases} \mathbf{V}_r^- = \mathbf{V}_a^- - \mathbf{V}_p \\ \mathbf{V}_r^+ = \mathbf{V}_a^+ - \mathbf{V}_p \end{cases} \tag{3}$$

Notice that in the above expression we supposed that the time of flight in SOI is significantly less than the total flight time of the spacecraft. Thus, we neglected it when calculated asymptotic velocities.

Since in our technique we consider only impulse-free gravity assists, further we assume that $V_r = |\mathbf{V}_r^-| = |\mathbf{V}_r^+|$.

The basic principle of a gravity assist maneuver is to turn the vector $\mathbf{V}_r$ of the asymptotic velocity of the spacecraft, so that the vector $\mathbf{V}_a$ of its heliocentric velocity is changed. This can be achieved by selecting the periapsis distance $r_\pi$ of the hyperbolic trajectory of the spacecraft and the inclination of this trajectory to some reference plane. Fig. 2 shows an example of a planar gravity maneuver, when the vector $\mathbf{V}_r$ turns in the plane formed by vectors $\mathbf{V}_p$ and $\mathbf{V}_r^-$. The turn occurs at an angle *α*, which is defined as follows: $\alpha = \arccos\frac{\mathbf{V}_r^- \cdot \mathbf{V}_r^+}{|\mathbf{V}_r^-||\mathbf{V}_r^+|}$, also the angle $\delta$ between $\mathbf{V}_r^-$ and $\mathbf{V}_\text{p}$ should be defined as $\delta = \arccos\frac{\mathbf{V}_p \cdot \mathbf{V}_r^-}{|\mathbf{V}_p||\mathbf{V}_r^-|}$.

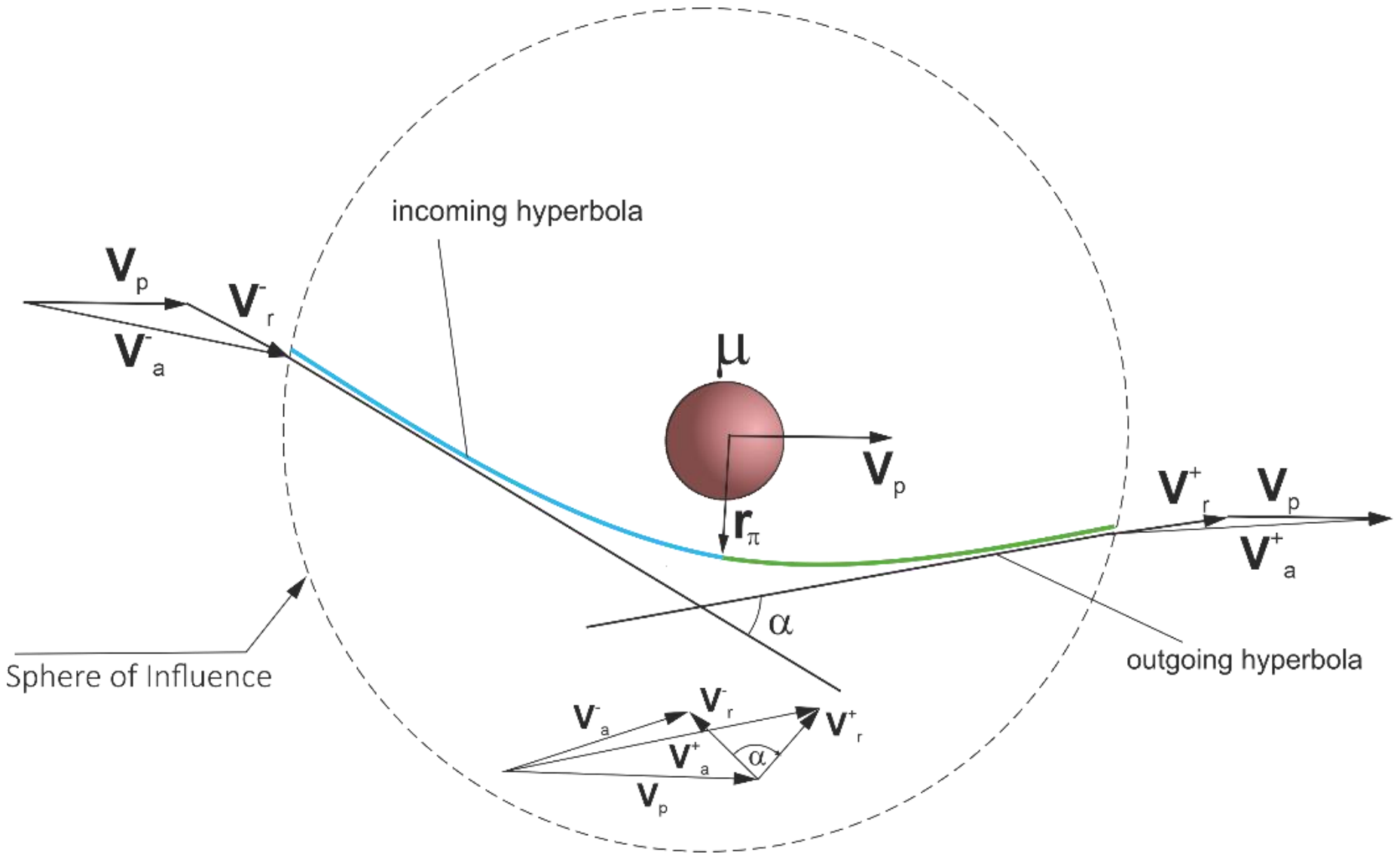

**Fig. 2.** Geometry of a gravity assist maneuver in the plane of the incoming trajectory.

In the presented method, the use of the gravity assist maneuver is required to change the parameters of the heliocentric trajectory of the spacecraft by approaching Venus, so that after its flyby the spacecraft transits into a *m*:*n* resonant orbit. However, unlike the case described above (the case of a planar gravity assist maneuver) here the principle of spatial maneuver is applied, when, at first, the vector $\mathbf{V}_r$ of asymptotic velocity turns in the plane formed by vectors $\mathbf{V}_r^-$ and $\mathbf{V}_p$ by angle $\alpha_{min}$ (see Fig. 3), and then it turns by angle *γ* in the plane orthogonal to the vector $\mathbf{V}_\text{p}$. Note that the angle between $\mathbf{V}_r^-$ and $\mathbf{V}_p$ remains the same for any *γ*. In this case the spacecraft can transfer between orbits of the same *m*:*n* resonance. In a heliocentric system this results in that at different *γ* the semi-major axis of the spacecraft's heliocentric orbit remains unchanged ($a = const$), while the eccentricity ($e$) and the inclination ($i$) change so that $\sqrt{1-e^2}\cos i = const$. To determine the required asymptotic velocity projections, it is convenient to use an approach (Strange, Russell, and Buffington 2008; Golubev, Grushevskii, Koryanov, et al. 2019b), in which it is proposed to describe a sphere of radius $V_r$ showing all possible directions of rotation of the incoming asymptotic velocity. However, not all directions are achievable, and the maximum natural turn angle *α** should be calculated in order to find the achievable ones. The value of *α** is determined by the minimum allowed radius of the periapsis $r_\pi$ for a given value of $V_r$, by the following relation:

$$\sin\frac{\alpha^*}{2} = \frac{1}{1 + r_{\pi,\min} V_r^2 / \mu}, \tag{4}$$

where $r_{\pi,\min}$ denotes the minimal periapsis distance. In this work the value $r_{\pi,\min} = 6551\ km$ was considered to prevent the spacecraft trajectory to be affected by Venusian atmosphere.

The $\alpha^*$ defines the segment of the sphere shown in Fig. 3, bounded by a circle located at an angular distance α* from the vector $\mathbf{V}_r^-$.

Notice that the circle formed by rotation of the end of the vector $\mathbf{V}_r^+$ around the vector $\mathbf{V}_p$, as shown in Fig. 3, is a set of positions of the end of the spacecraft velocity vector $\mathbf{V}_a^+$ in the heliocentric reference frame frozen at the moment of the maneuver and the vector $\mathbf{V}_r^+$ of the spacecraft outgoing velocity, for cases when after the gravity assist maneuver heliocentric periods of orbits of Venus and the spacecraft are in the *m:n* ratio.

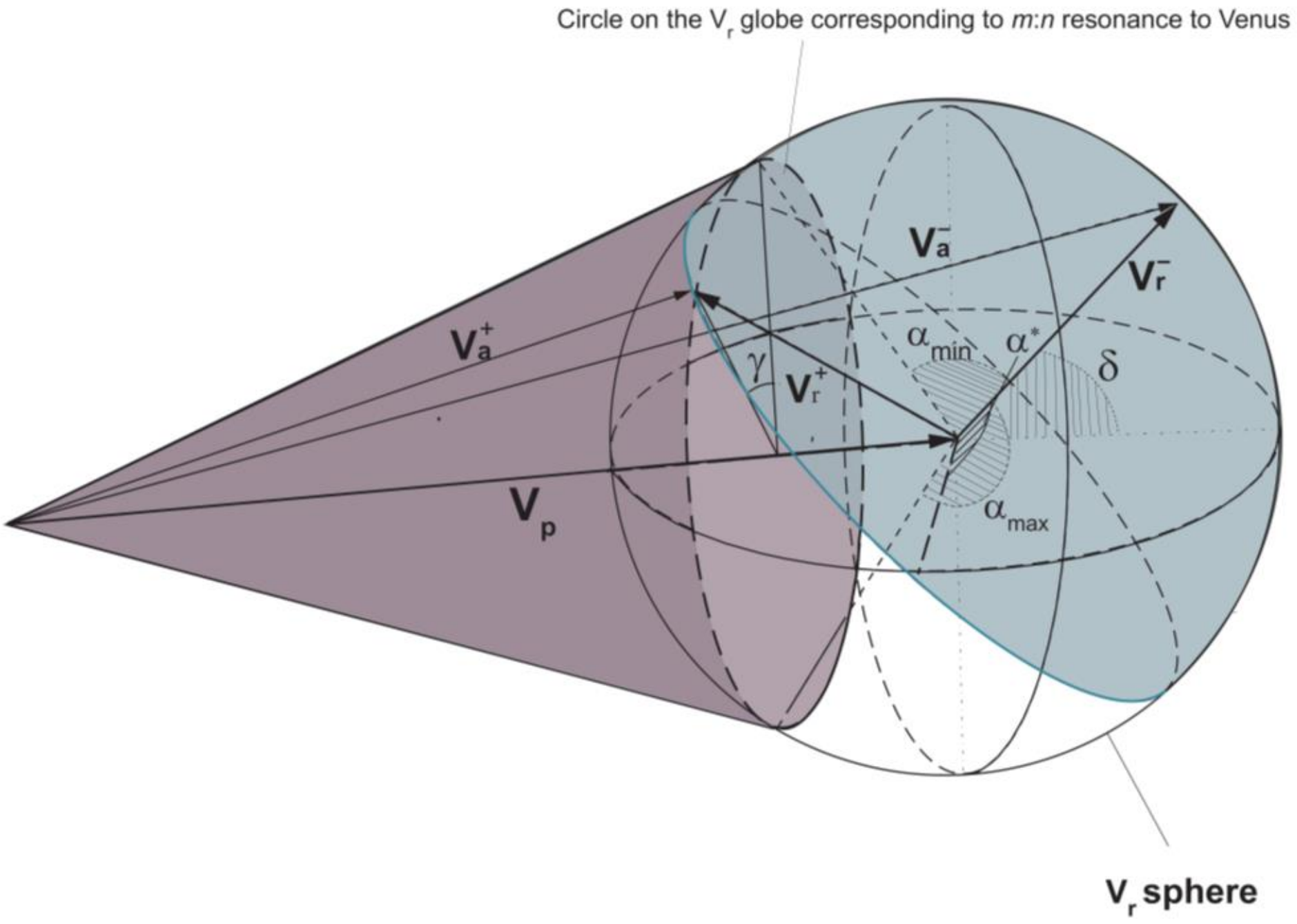

**Fig. 3.** Geometry of a gravity assist maneuver with the transfer of the spacecraft to a *m*:*n* resonant orbit.

In order for the spacecraft orbit to be resonant to the planet one, the following condition should be met:

$$|V_p - V_r| \le V_a^+ \le V_p + V_r\,. \tag{5}$$

According to it, for at least one resonant orbit to exist, it is necessary that $V_a^+$ is achievable at the maximum turn of the vector of the incoming asymptotic velocity against the velocity vector of the planet (the left part in (5)) or along the vector of velocity the planet (the right part in (5)). However, this condition is only necessary, but not sufficient.

For the existence of at least one solution it is required that the following condition is met:

$$V_r^- = V_r^+,\ \alpha^* > \alpha_{\min}\,. \tag{6}$$

Proofs of the statements made that the above conditions (5) and (6) actually determine the existence of at least one feasible solution in which the spacecraft can achieve the required landing point using a gravity assist maneuver and a resonant orbit are given in **Appendix A**.

The above mentioned inequalitites can be expanded to the following three cases:

$$\begin{cases} \alpha_{\min} \le \alpha_{\max} \le \alpha^*, & (7.1) \\ \alpha_{\min} \le \alpha^* \le \alpha_{\max}, & (7.2) \\ \alpha^* \le \alpha_{\min} \le \alpha_{\max}. & (7.3) \end{cases}$$

The first case (7.1) is the most favorable because all resonant orbits of the chosen resonance *m*:*n* are accessible, in this case the maximal number of landing points are attainable. The second case (7.2) shows that some of the resonant orbits are inaccessible and thus some points on the Venus surface can not be attainable. In the last case (7.3) the impulse-free transfer to a resonant orbit by gravity assist maneuver at Venus is not possible, i.e. there is no feasible solution exists.

Minimum and maximum required turn angles $\alpha_{min}$ and $\alpha_{max}$ can be defined from the following equations (see A.f, considering the spacecraft velocity in a resonant orbit $V_{res} = V_p \sqrt{2 - \left(\frac{m}{n}\right)^{-2/3}}$ and assuming that the orbit of the planet is circular):

$$\alpha_{\min} = \arccos\left[\frac{1}{2V_r V_p}\left\{\left(-\left(\frac{m}{n}\right)^{\frac{2}{3}} + 1\right)V_p^2 - V_r^2\right\}\right] - \delta,$$

$$\alpha_{\max} = \arccos\left[\frac{1}{2V_r V_p}\left\{\left(-\left(\frac{m}{n}\right)^{\frac{2}{3}} + 1\right)V_p^2 - V_r^2\right\}\right] + \delta.$$

It should be noted that the angle $\alpha$ varies from $\alpha_{min}$ to $\alpha_{max}$.

### 1.3. Analysis of possible *m*:*n* resonances and its feasibility to application in Venus exploration mission scenario

Let us analyze a possible resonant orbit where the spacecraft can be transferred by an impulse-free Venus gravity assist.

By converting the expression (5) and considering the connection of the large semi-axes of orbits of the spacecraft and the planet by virtue of Kepler's third law, we may rewrite the necessary condition for the existence of the required resonance in the form:

$$\sqrt{\frac{{\mu_{Sun}}^3}{\left(2\mu_{Sun} - r_p\left[\left(V_r^+\right)^2 + V_p^2 + 2V_r^+V_p\right)\right]^3}} \le \frac{m}{n} \le \sqrt{\frac{{\mu_{Sun}}^3}{\left(2\mu_{Sun} - r_p\left[\left(V_r^+\right)^2 + V_p^2 - 2V_r^+V_p\right)\right]^3}}.$$

Let us construct the maximum permissible *m*:*n* ratios of periods of the spacecraft and the planet depending on the asymptotic velocity of the spacecraft during the first flyby of Venus. Fig. 4 shows the $\alpha^*$ vs. $V_r$, which displays that with increasing $V_r$ the available relation *m*/*n* grows as well. However, the apparent growth of the available resonances turns out to be limited by the decreasing a value of $\alpha^*$ with increasing $V_r$.

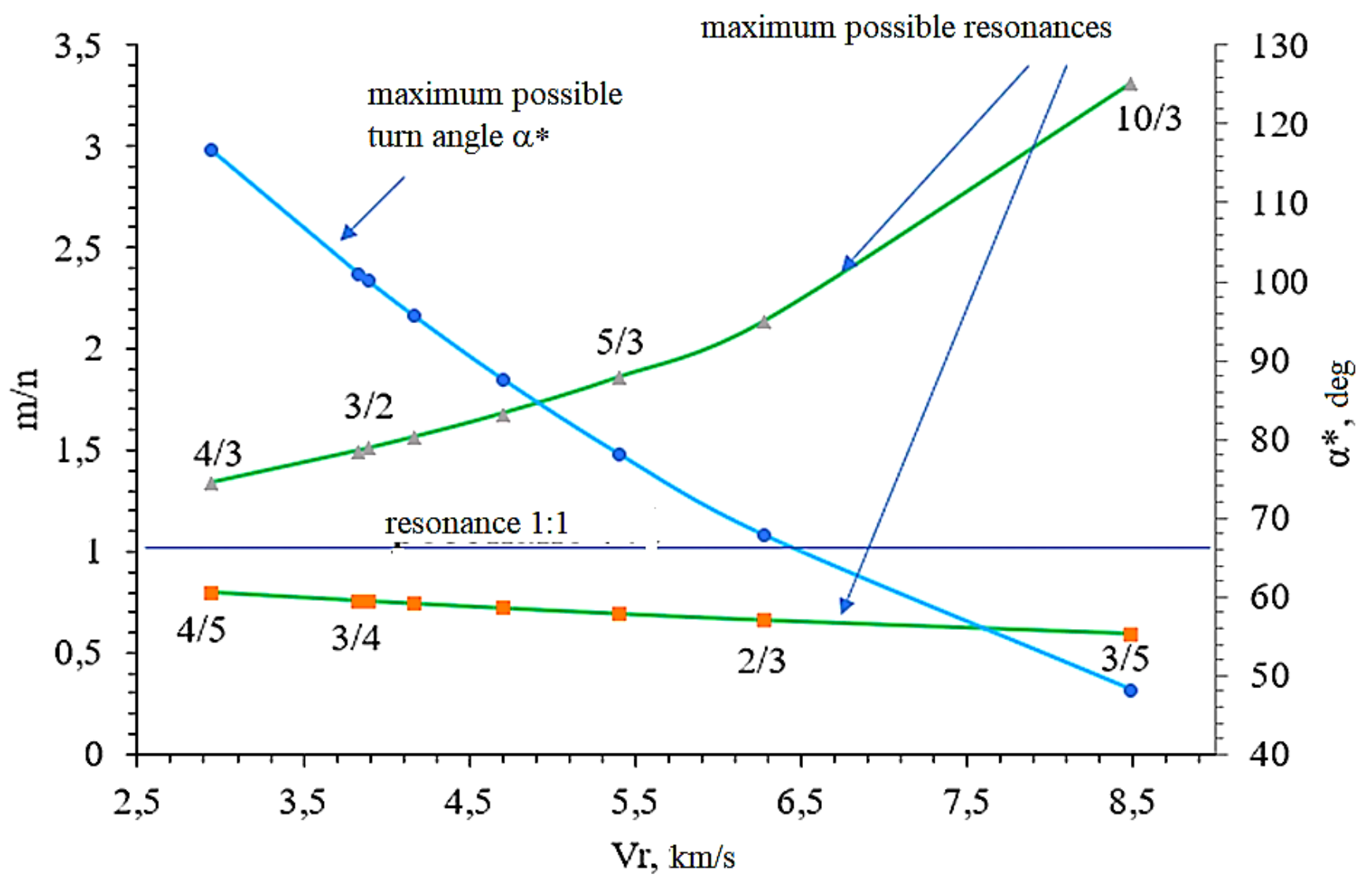

**Fig. 4.** The maximum possible resonant ratios of orbits of the spacecraft and Venus after the gravity assist maneuver

Fig.4 shows that at low values of $V_r$ (from 2.7 to 3.5 km/s), possible resonance ratios range from 4:3 to 3:4, while $\alpha^*$ reaches its maximum value at this interval of asymptotic velocities. At the same time, $\alpha^*$ decreases with an increase in the asymptotic velocity. From this we can conclude that in order to satisfy the condition (7.1), which actually characterizes the effectiveness of our method, it is required to approach Venus with the lowest possible value of the asymptotic velocity.

Let us denote the following parameters:

$$\gamma_{\max} = \begin{cases} \arccos\left(\dfrac{\cos\alpha^* - \cos(\alpha_{\min}+\delta)\cos\delta}{\sin(\alpha_{\min}+\delta)\sin\delta}\right), & \cos(\alpha_{\max}) \le \cos\alpha^* \le \cos(\alpha_{\min}) \\ \pi, & \cos\alpha^* \le \cos(\alpha_{\max}) \le \cos(\alpha_{\min}) \\ \textit{no solution exists}, & \cos\alpha^* \ge \cos(\alpha_{\min}) \end{cases},$$

and

$$\Delta\gamma = 2\gamma_{\max},$$

where $\gamma_{max}$ is the maximum value of the angle $\gamma$ by which the projection of the vector $\mathbf{V}_r^-$ of the incoming asymptotic velocity of the spacecraft can be rotated in the plane orthogonal to the $\mathbf{V}_p$ vector in order to enter an orbit with the $m$:$n$ resonance ratio. The $\Delta\gamma$ angle, in turn, characterizes the total amount of available directions of turn of the $\mathbf{V}_r^-$ vector made by the gravitational field of Venus. Note that the $\gamma_{\max}$ exists in the case (7.2). Obviously, in the case (7.1) we obtain $\gamma_{\max} = \pi$, but $\Delta\gamma = 2\pi$, i.e., directions of the turn are allowed by $\alpha^*$. Fig. 5 shows the dependence of the $\Delta\gamma$ on the value of the asymptotic velocity of the spacecraft and the $\delta$ angle characterizing the position of $\mathbf{V}_r^-$ relative to $\mathbf{V}_p$.

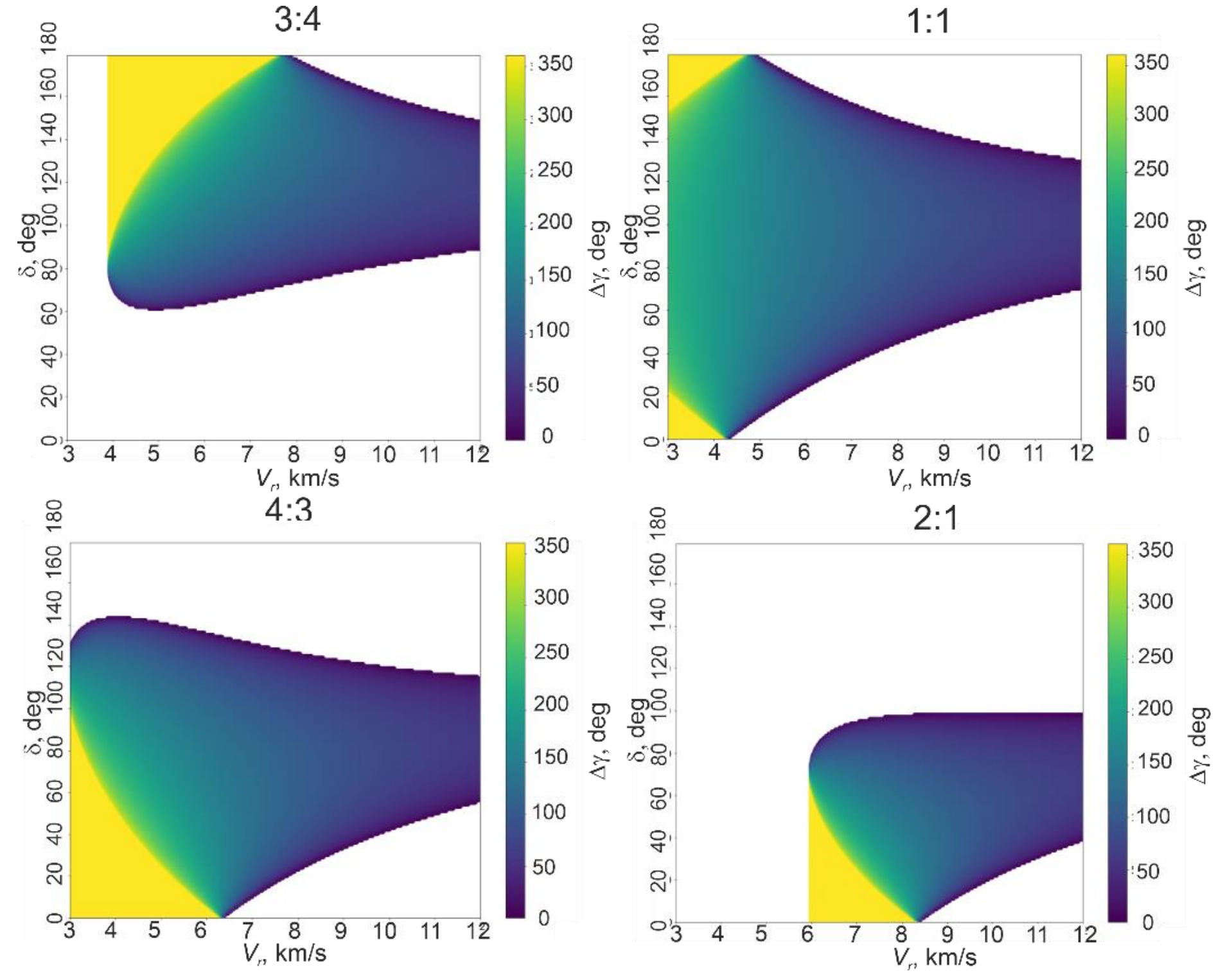

**Fig. 5.** Dependence of $\Delta\gamma$ on $V_r$ and δ for the 3:4, 1:1, 4:3 and 2:1 resonant ratios. Forbidden regions, where the resonance is not achieved for a given asymptotic velocity of the spacecraft and its orientation, are shown in white.

Note that, according to the dependencies shown in Fig. 5, the 3:4 and 4:3 resonant ratios are characterized by rather large ranges of $V_r$ and $\delta$, in which the condition (7.1) is fulfilled for any vector of the asymptotic velocity of the spacecraft when it is rotated to the line of a given resonance (an example of such a line is given in Fig. 3 for the *m:n* resonance). In turn, the 2:1 resonance ratio which is realized when the spacecraft moves along a heliocentric orbit with a high aphelion cannot be used in principle at a low asymptotic velocity of the spacecraft. The 1:1 resonance ratio is realized over a wide range of $V_r$ and $\delta$, however, the areas where Δγ=360 deg. are smaller compared to the cases of the 3:4 and 4:3 resonances. On the other hand, due to the limited resource of the scientific instruments and service systems of the spacecraft, it is often desirable to reduce the flight duration, i.e. to ensure the fastest meeting of the spacecraft with Venus, while maintaining the condition of an impulse-free Venus flyby. Thus, it is obvious that the spacecraft motion along the 1:1 resonant orbit is the best strategy in terms of the fastest encounter with Venus as well as the compromise in $\Delta\gamma$ magnitude. Therefore, further in the work, all examples will be given specifically for 1:1 resonant orbits.

Note that the parameters $\alpha_{\min}, \alpha_{\max}$ defined above are simplified for the case of the 1:1 resonance ratio and can be written as:

$$\alpha_{\min} = -\delta + \arccos\left(-\frac{V_r}{2V_p}\right),$$

$$\alpha_{\max} = \delta + \arccos\left(-\frac{V_r}{2V_p}\right).$$

## 1.4. Selection of the required resonant orbit that ensures landing in a given area on the surface of Venus

This subsection discusses the sequence of actions connecting the given landing point on the surface of Venus with the required resonant heliocentric orbit that ensures landing at this point. It is assumed that the following parameters are known (see Nomenclature): $\lambda_L$ and $\varphi_L$; $t_1$; $t_k$; $\mathbf{r}_p$ and $\mathbf{V}_p$, as well as the Venus orientation parameters at the specified time moments.

First of all, the transition is made from the inertial ecliptic system, in which the parameters of the motion of Venus and the spacecraft are known, to the local orbital system $\xi\eta\zeta$ (see Fig. 6), the primary axis $\xi$ is directed along the vector of orbital velocity of Venus, the secondary axis $\zeta$ is directed along the vector of the orbital angular momentum of Venus and the axis $\eta$ completes the system. The following unit vectors are fully determinate the local orbital system:

$$\xi^0 = \frac{\mathbf{V}_p}{|\mathbf{V}_p|}, \quad \zeta^0 = \frac{\mathbf{r}_p \times \mathbf{V}_p}{|\mathbf{r}_p \times \mathbf{V}_p|}, \quad \eta^0 = \xi^0 \times \zeta^0.$$

Origin of the chosen coordinate system is placed at the end of the Venus heliocentric velocity vector.

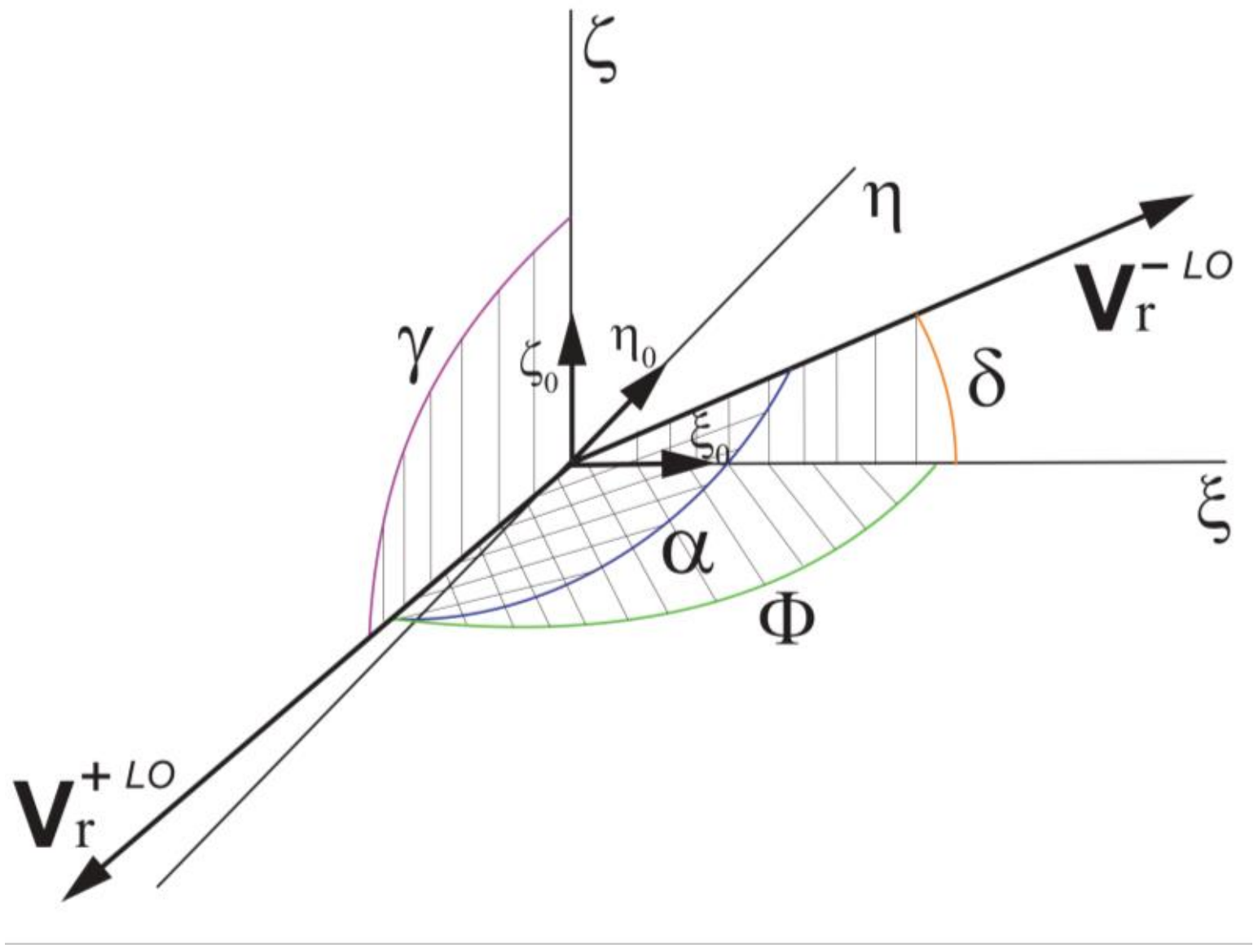

**Fig. 6.** Orientation of the $\mathbf{V}_r^{+LO}$ and $\mathbf{V}_r^{-LO}$ vectors in the local orbital system.

At the first step $\mathbf{V}_p$ and $\mathbf{V}_r^-$ should be transferred from the ecliptic system to the local orbital one. Notice that for description of vectors in this system we use the upper script *LO (i.e. local orbital)*. The $\mathbf{V}_r^{+LO}$ vector in the local orbital system (see Fig. 6) is calculated as follows:

$$\mathbf{V}_r^{+LO} = V_r \{\cos\Phi,\ \sin\gamma\sin\Phi,\ \cos\gamma\sin\Phi\}^T, \tag{8}$$

where $\Phi = \arccos\left[\frac{1}{2V_rV_p}\left\{\left(-\left(\frac{m}{n}\right)^{\frac{2}{3}}+1\right)V_p^2 - V_r^2\right\}\right]$ is the angle between $\mathbf{V}_r^+$ and $\mathbf{V}_p$. To determine the required vector of the asymptotic velocity of the spacecraft, it is necessary to find the specific $\gamma$ angle related to the given landing point. Using the relations of spherical geometry, the following expression can be obtained to calculate this angle:

$$\gamma_{1,2} = \gamma_C \pm \arccos\frac{\cos\psi}{\cos|\Phi - \sigma|}, \quad (9)$$

where $\gamma_{1,2}$ corresponds to two possible solutions for "+" and "-" cases in (9), $\gamma_C = arctg\left(-\frac{\sin(\lambda_L - \lambda_P)\cos\varphi_L}{\sin\varphi_L\cos\varphi_P - \cos\varphi_L\sin\varphi_P\cos(\lambda_L - \lambda_p)}\right)$ is defined by known the latitudes and the longitudes of the landing point $L(\varphi_L, \lambda_L)$ and $(P(\varphi_P, \lambda_P))$ using the law of cosines, $\sigma = \frac{\pi}{2} - \arcsin\left(\sin\varphi_L\sin\varphi_P + \cos\varphi_L\cos\varphi_P\cos(\lambda_L - \lambda_p)\right)$ is the angular distance between $L(\varphi_L, \lambda_L)$ and $P(\varphi_P, \lambda_P)$. Notice that planetographic coordinates are given at the moment $t_k$, i.e. at the landing time.

Substituting (9) into (8) and considering the expression for $\gamma_C$, we may find the vector of the outgoing asymptotic velocity of the spacecraft after the turn:

$$\mathbf{V}_r^{+LO} = V_r \begin{Bmatrix} \cos\Phi \\ \sin\left(arctg\left(-\frac{\sin(\lambda_L - \lambda_P)\cos\varphi_L}{\sin\varphi_L\cos\varphi_P - \cos\varphi_L\sin\varphi_P\cos(\lambda_L - \lambda_p)}\right) \pm \arccos\frac{\cos\psi}{\cos|\Phi - \sigma|}\right)\sin\Phi \\ \cos\left(arctg\left(-\frac{\sin(\lambda_L - \lambda_P)\cos\varphi_L}{\sin\varphi_L\cos\varphi_P - \cos\varphi_L\sin\varphi_P\cos(\lambda_L - \lambda_p)}\right) \pm \arccos\frac{\cos\psi}{\cos|\Phi - \sigma|}\right)\sin\Phi \end{Bmatrix}.$$

In this case, the resulting velocity vector must be achievable due to a gravity assist maneuver, i.e. the condition $\alpha \le \alpha^*$ should be met.

If the above condition is violated, then the flight in such an orbit cannot be carried out, in this case a change is required either of the launch date or the date of arrival to Venus. Another option is to apply an impulse $\Delta V_n$ on the boundary of the planet's SOI that will give the additional turn to $\mathbf{V}_r^-$. The required characteristic velocity consumption can be calculated as follows:

$$\Delta V_n = 2V_r \sin\frac{\Delta\alpha}{2},$$

where $\Delta\alpha = \alpha^* - \alpha$. Let us estimate the minimum value of the impulse required to turn $\mathbf{V}_r^-$ by 1 deg if $V_r = |\mathbf{V}_r^-| = 2.9\ km/s$. The impulse magnitude calculated according to the equation above is $\Delta V_n = 101\ m/s$. Shifting dates of launch or arrival to avoid unnecessary characteristic velocity consumption seems to be the most preferable, thus below we consider only such trajectories that do not require any additional $\Delta V$ during the Venus flyby. If the reachability condition is fulfilled, then operations with the $\mathbf{V}_r^{+LO}$ vector continue.

The further operations with vectors are performed in the ecliptic coordinate system. As a result of the presented algorithm, the state vector of the spacecraft at the moment $t_1$ is completely determined by the system of equations

$$\begin{cases} \mathbf{V}_a^+(t_1) = \mathbf{V}_r^+(t_1) + \mathbf{V}_p(t_1) \\ \mathbf{r}(t_1) = \mathbf{r}_p(t_1) \end{cases},$$

*Remark 1*. From the above relation for the $\mathbf{V}_r^{+LO}$ vector it can be seen that there are two solutions, i.e. two resonant orbits leading the spacecraft to the same landing point, exist. The main difference between them lies in the height of the periapsis of the planetocentric trajectory of the spacecraft, as well as in the inclination of the plane of this trajectory to the Venus equator plane.

*Remark 2*. Among two possible trajectories leading to the required landing point, any trajectory can be selected if one of the conditions (7.1)-(7.3) is met. The choice of a specific $\mathbf{V}_r^{+LO}$ vector is determined by additional restrictions on the ballistic scenario of the mission to Venus.

*Remark 3*. The required landing point comes to its position planned for landing not when the first Venus flyby is performed, but after a Venusian year (224.7 days). Notice that at first glance, it may seem too long to fly for about 1 Venusian year and is too high a price to pay for landing in the required region. However, if we compare the proposed approach with the traditional ones, which consist, for example, of shifting the launch date (which will eventually increase $\Delta V_\Sigma$) or landing from elliptical or circular orbit (which, of course, requires a very high characteristic velocity consumption to break into the near-Venusian orbit), the cost in the discussed case would not be so high. Since in our method and because of the possibility to control the asymptotic velocity vector, the choice of the landing site is practically unlimited.

### 1.5. Optimization criteria

Total costs $\Delta V_\Sigma$ of the characteristic velocity required for the flight along the Earth-Venus-Venus trajectory may be considered as a sum of an impulse $\Delta V_0 = \Delta V_0(t_0, t_1)$ applied to the spacecraft at the moment $t_0$ of launch from the low Earth orbit[6] and an impulse $\Delta V_1 = \Delta V_1(t_0, t_1, t_k)$ applied at the moment $t_1$ at the 1st flyby trajectory periapsis:

$$\Delta V_\Sigma(t_0, t_1, t_k) = \Delta V_0(t_0, t_1) + \Delta V_1(t_0, t_1, t_k).$$

Since in the presented method only an impulse-free transition of the spacecraft to Venus-Venus arcs is considered, we may set $\Delta V_1=0$. This condition serves as a filter to the optimized trajectories: if it is violated then flight along such trajectory can not be made as well as landing in the selected point.

Thus, the problem of optimization the flight trajectory with a flyby of Venus is reduced to the problem of finding the minimal cost of the launch characteristic velocity, and the optimization criterion in this case can be written as:

$$F(t_0, t_1) = \Delta V_\Sigma(t_0, t_1) = \Delta V_0(t_0, t_1) \to \min \quad . \tag{10}$$

Notice that in the presented method we did not consider any other impulses at Earth-Venus-Venus arcs. It means that in general case the criterion (10) does not correspond to the optimal solution. For example, if an additional impulse is applied at the Earth-Venus arc, then such solution may consume less $\Delta V_\Sigma$ than the trajectory optimized only by the criterion (10). Also, the more generalized options are described in works (Casalino, Colasurdo, and Pastrone 1998; Carletta 2021). In our model in order to simplify the search of trajectories, we consider only the case when no additional impulses along the Earth-Venus and the Venus-Venus arcs are applied. However, in general case the similar analysis can also be made using the presented method.

Since our goal is to provide an access to as many landing sites on the surface of Venus as possible, i.e. to fulfill the condition (3) on as many trajectories as possible, we should modify the optimization criterion (10) by adding the asymptotic velocity $V_r(t_0, t_1)$ of the spacecraft during the 1st Venus flyby and consider the criterion in the following form:

$$\tilde{F}(t_0, t_1) = \Delta V_0(t_0, t_1) + V_r(t_0, t_1) \to \min . \quad (11)$$

[6] Parameters of the low Earth orbit are: periapsis distance ($r_{\pi 0}$), 6571 km, inclination to the Earth equator (J2000) ($i_0$) 51.6 deg.

Thus, the optimization problem is to find the minimum of $\tilde{F}(t_0, t_1)$ on the set of all possible values of $t_0$ and $t_1$.

Note that the use of the criterion (11) allows one to increase a number of areas attainable for landing on the surface of Venus due to conditions (7.1)-(7.3) since the minimization of $V_r$ along with $\Delta V_0$ leads to increase the natural turn angle and thus provide more options for an impulse-free flyby of Venus and eventually increase a number of attainable landing areas.

# 2. Results

## 2.1. Determination of launch windows to Venus

In Table 1 the results of our calculation of launch dates in 2029-2037 optimized by the criterion (11) using the Broyden–Fletcher–Goldfarb–Shanno algorithm with the preliminary search of solutions by the differential evolution algorithm are presented. The table contains optimal dates of launch and landing as well as some other parameters ($\Delta V_0$, $V_r$) of optimal trajectories for the considered years. Notice that in this paper trajectories in which the Earth-Venus arc has the angular distance less than 180 deg are designated as the 1st semi-turn trajectories, while trajectories for those the angular distance is more than 180 deg are called 2nd semi-turn ones.

**Table 1.** Optimal launch dates, time of flight and $\Delta V_0$, $V_r$ parameters for the 1st semi-turn and the 2nd semi-turn trajectories to Venus

| 1st semi-turn | | | | 2nd semi-turn | | | |
|---|---|---|---|---|---|---|---|
| Optimal launch date | Time of flight, days | $\Delta V_0$, km/s | $V_r$, km/s | Optimal launch date | Time of flight, days | $\Delta V_0$, km/s | $V_r$, km/s |
| Nov 14, 2029 | 108.0 | 4.12 | 3.28 | Oct 22, 2029 | 159.7 | 3.58 | 4.82 |
| Jun 05, 2031 | 127.0 | 3.79 | 2.90 | May 22, 2031 | 158.0 | 3.52 | 3.81 |
| Dec 21, 2032 | 135.0 | 3.55 | 3.52 | Dec 07, 2032 | 158.2 | 3.70 | 2.66 |
| Aug 17, 2034 | 110.8 | 3.70 | 4.50 | Jun 09, 2034 | 181.3 | 3.86 | 2.97 |
| Apr 05, 2036 | 104.9 | 4.19 | 4.15 | Dec 23, 2035 | 199.7 | 4.29 | 3.15 |
| Nov 12, 2037 | 105.2 | 4.13 | 3.33 | Oct 17, 2037 | 164.1 | 3.60 | 4.89 |

## 2.2. Design of a trajectory to Venus with landing in the Vellamo-South region

Consider one of possible landing sites on the Venus surface: the Vellamo-South region (29°S, 164°E) located in the southern part of the Vellamo Planita vast plain, which is important from a scientific point of view and satisfies the landing safety criterion. This region is of scientific importance because, according to (Ivanov, Zasova, and Gregg 2018; Ivanov, Zasova, Zeleny, et al. 2017; Ivanov et al. 2021; Ivanov 2016; Basilevsky et al. 2007), the Vellamo Planita is almost completely overlain by sediments of the lower subdivision of the regional plains. The flat part stretches for 300 km, there is the longest lava channel on Venus Baltic Vallis (Ivanov, Zasova, and Gregg 2018; Ivanov et al. 2021).

An example of design a trajectory with landing in Vellamo-South at launch in the May 20, 2031 - June 18, 2031 launch window is presented below. The optimal launch date calculated by trajectory optimization under the criterion (11) was defined as June 05, 2031; the total cost of the characteristic velocity is $\Delta V_\Sigma$=3.79 km/s. Note that at launch in 2031 the Vellamo-South region stays unattainable using the standard flight trajectory without applying a gravity assist and transition to a resonant orbit (see Fig. 7).

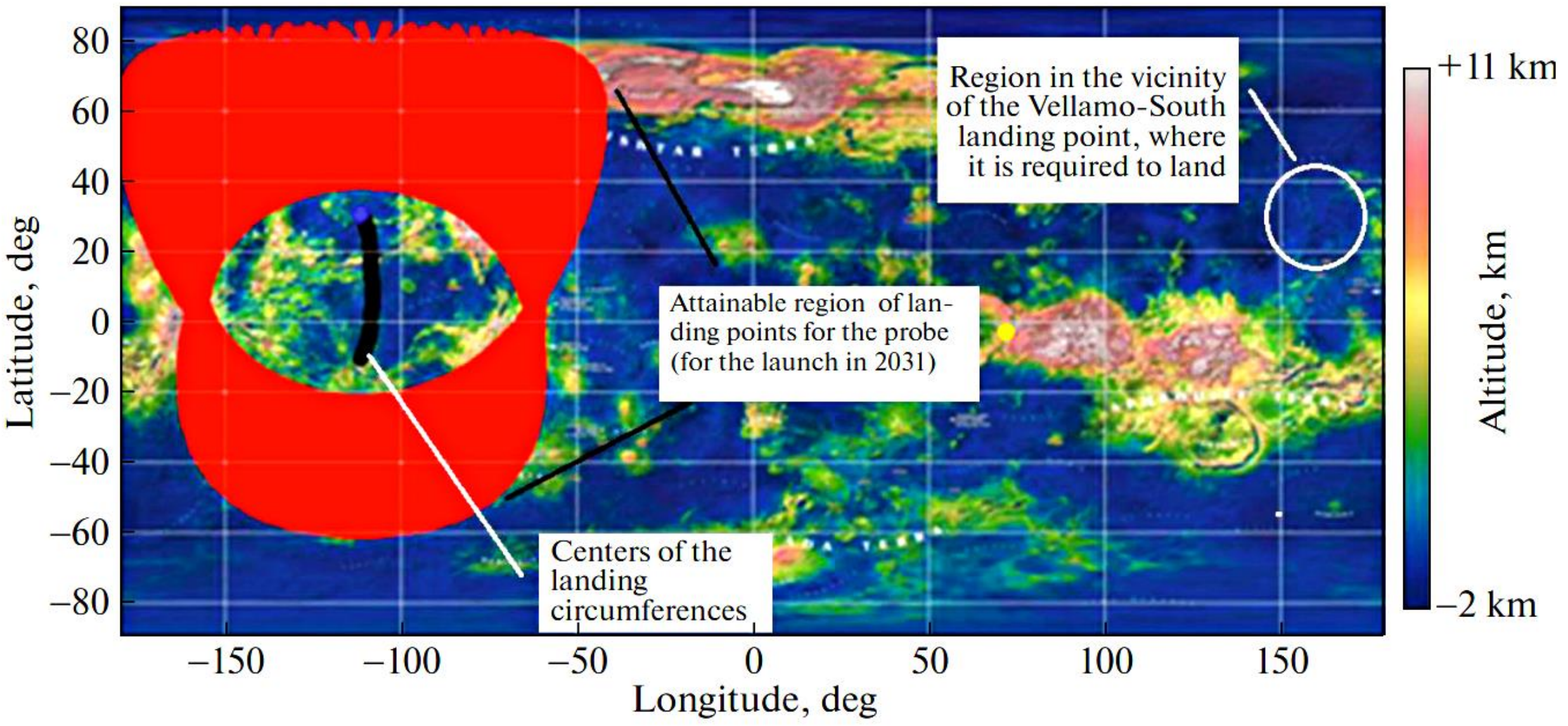

**Fig. 7.** Landing points for the flight to Venus at launch in the extended launch date interval May 20, 2031-Jun 18, 2031 using the direct flight trajectory. Red dots correspond to the attainable areas on the surface of Venus, black and blue dots indicate centers of landing circles. Dates on the figures are given in the format dd.mm.yyyy.

As an example, we consider launch at June 03, 2031, since this date belongs to the launch window and is close to the optimal one according to the criterion (11); the $\Delta V_0$ value at launch on this date is 3.78 km/s. The flight time is defined by the optimization problem solved considering the criterion (11). The angle of entry into the atmosphere is assumed to be 12 degrees, which approximately corresponds to a maximum overload of 100g[7]. In accordance with the conclusions obtained above, we consider only 1:1 resonant orbits, despite the fact that the algorithm allows, in principle, to vary and choose any resonance ratio of the orbits of the spacecraft and Venus realized by an impulse-free gravity assist maneuver.

Table 2 presents characteristics of two different 1:1 resonant orbits providing landing in the Vellamo-South region.

**Table 2.** Some characteristics of the trajectories to Venus including flight on a 1:1 resonant orbit and providing landing in the Vellamo-South region

| | Celestial body (launch, flyby, landing) | Date of launch/flyby/landing | Height of the periapsis of the spacecraft trajectory moving in the SOI of the celestial body, km | Spacecraft asymptotic velocity at the celestial body, km/s | Cost of the launch/flyby $\Delta V_0$, km/s |
|---|---|---|---|---|---|
| $i$=2.1 deg | Earth | Jun 03, 2031 | 200 | 3.56 | 3.78 |
| | Venus | Oct 08, 2031 | 13233 | 2.91(31.9°, 73.8 °)* | 0** |

[7] Notice that 100g is approximate and estimated using equation in e.g. (Eismont et al. 2021a). Such overload probe withstands during the descent at an altitude about 70-80 km.

| | | | | | |
|---|---|---|---|---|---|
| | Venus (Vellamo-South) | May 20, 2032 | - | 2.91 | 0** |
| $i$=4.21 deg | Earth | Jun 03, 2031 | 200 | 3.56 | 3.78 |
| | Venus | Oct 08, 2031 | 6573 | 2.91(49.8$^{\circ}$, -12.5$^{\circ}$)* | 0** |
| | Venus (Vellamo-South) | May 20, 2032 | - | 2.91 | 0** |

Notes: * –orientation of the asymptotic velocity vector after turn by the gravitational field of Venus in the Venusian ecliptic coordinate system in the direct ascension and the declination coordinates is given; ** - the cost of correcting the trajectory of the spacecraft during the Venus flyby, as well as corrections of the interplanetary trajectory are not taken into account and are not considered in this method.

Note that when flying to Venus at launch on June 03, 2031, the condition (7.1) is fulfilled for any of the resonant vectors of the asymptotic velocity. We show the change in the inclination of the possible 1:1 resonant orbits depending on the $\gamma$ angle (see Fig. 8). In the same figure, we show the radii of the periapsis at Venus of the spacecraft flight trajectories.

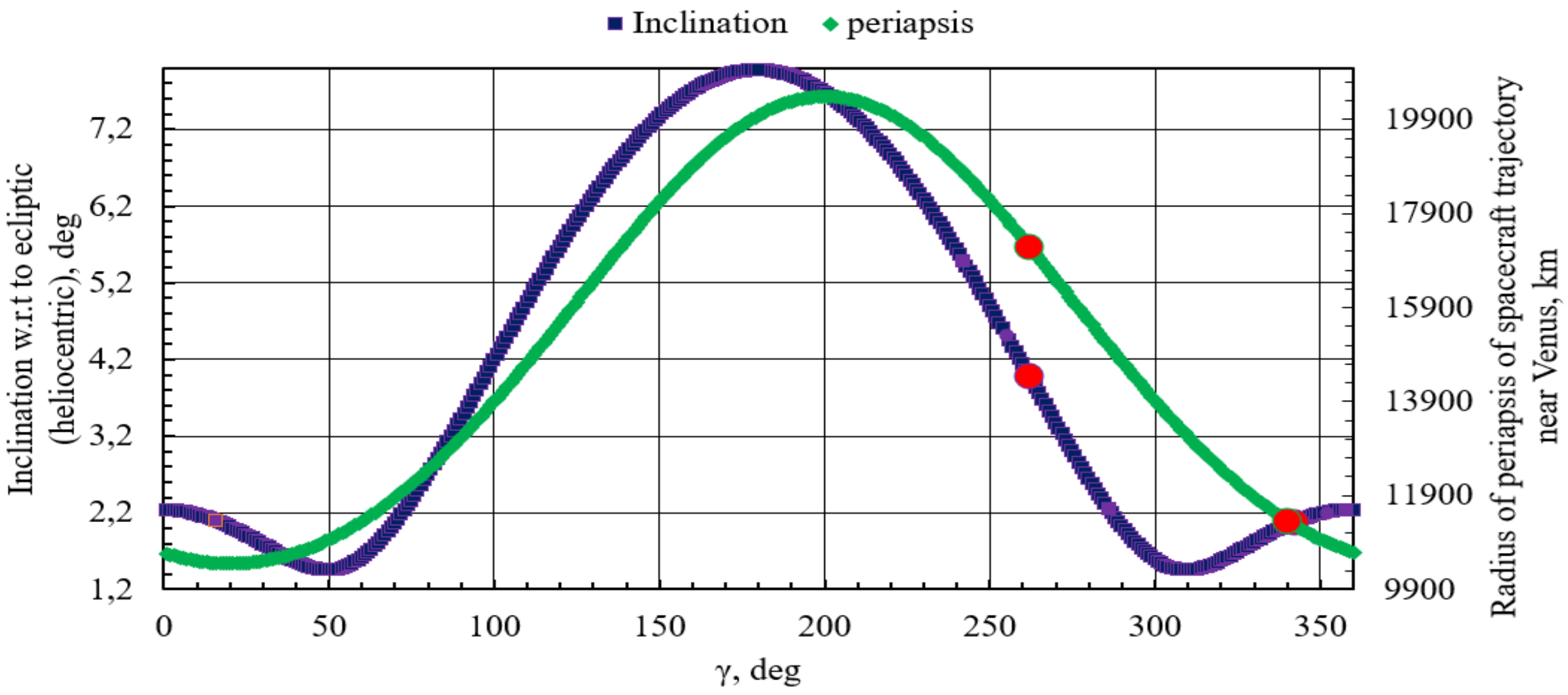

**Fig. 8.** Dependence of the ecliptic inclination for 1:1 resonant orbits and the radius of the periapsis of the spacecraft trajectory at the Venus flyby on the $\gamma$ angle. The red dots show the inclinations and the periapsis radii corresponding to the orbits leading to landing in the Vellamo-South region.

Note that the dependence of the inclination (see Fig. 8) on the $\gamma$ angle has two local maxima: when maneuvering in the vicinity of the North Pole of Venus, the maximum inclination is approximately 2.25 degrees; when maneuvering at the South Pole, the inclination of the heliocentric resonance orbit is approximately 8 degrees. The radii of the periapsis vary from 10 thousand to 20 thousand km, so as during the Venus flyby the spacecraft do not pass through the upper layers of its atmosphere.

Fig. 9 shows the heliocentric spacecraft trajectory including the Venus flyby and landing in the Vellamo-South region.

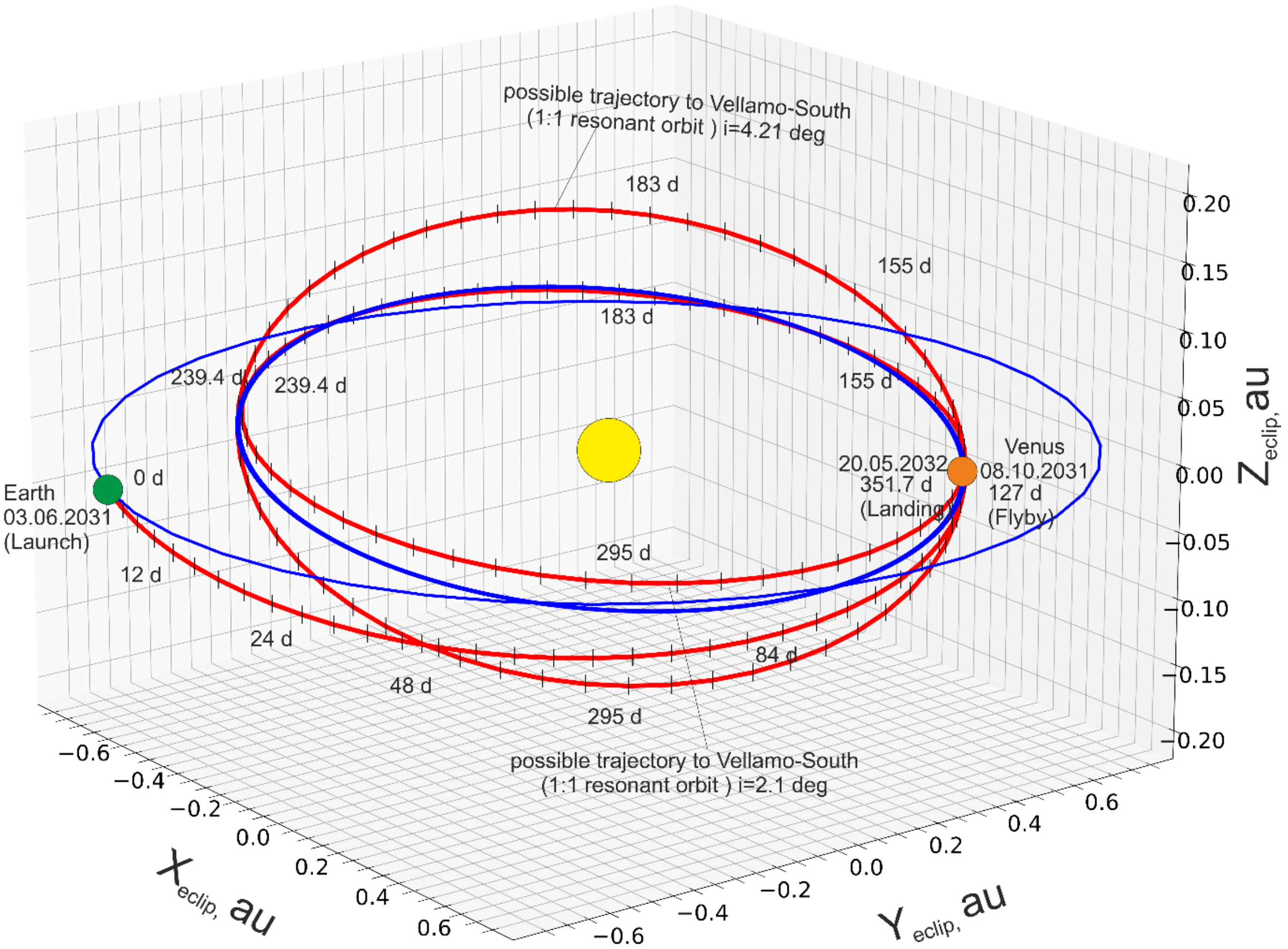

**Fig. 9.** The spacecraft trajectory to Venus including Venus flyby, the flight on the 1:1 resonant orbit and landing in the Vellamo-South region. Time intervals are marked by sticks corresponding to 1 day of flight. ($X_{Eclip}$,$Y_{Eclip}$,$Z_{Eclip}$) is the heliocentric ecliptic coordinate system.

Fig. 10 presents a map of the Venus surface with landing circles, obtained by the flight on the 1:1 resonant orbits leading to landing in the Vellamo-South region.

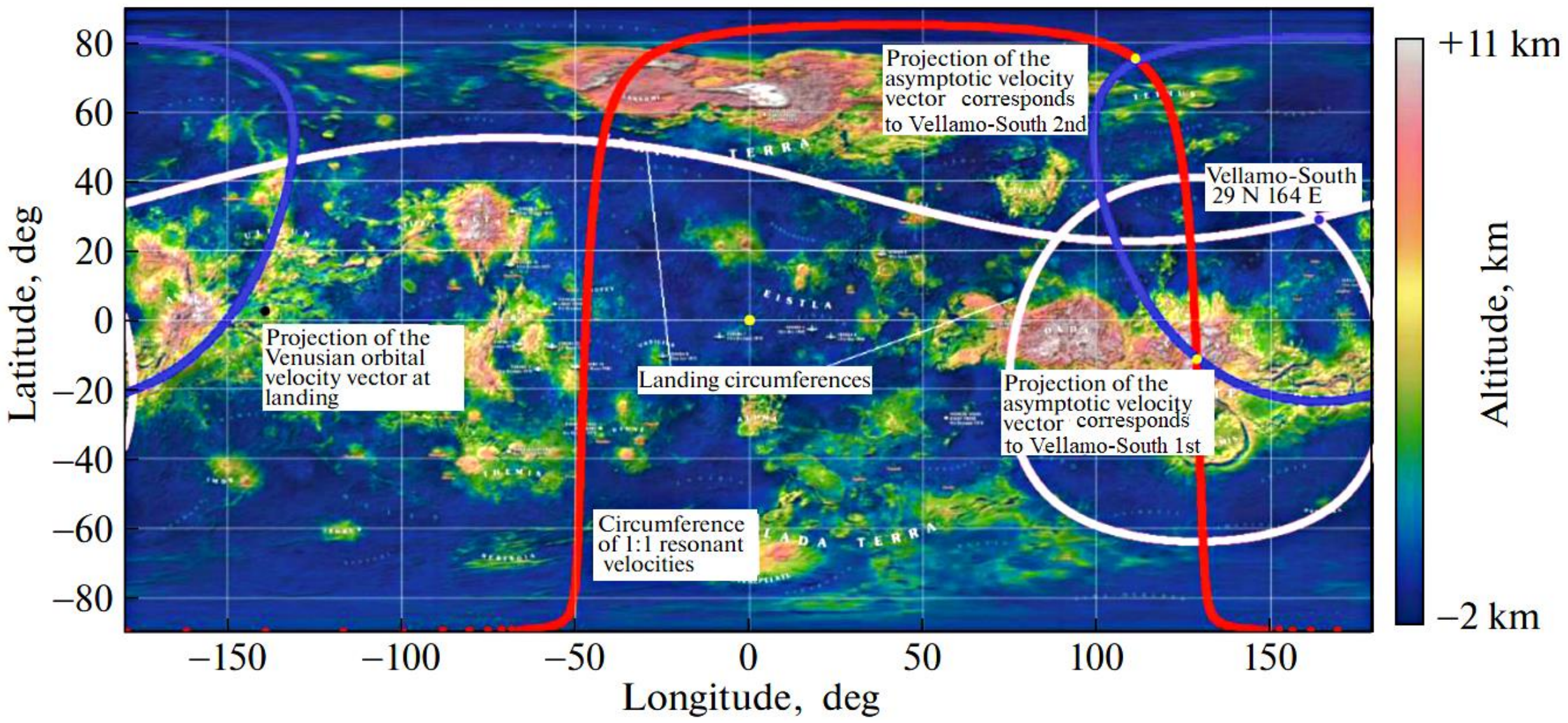

**Fig. 10.** Landing circles obtained during the flight along selected trajectories leading to landing in the Vellamo-South region. The blue circle is shown to highlight two points on the circle of resonant velocities that correspond to the required vectors of the asymptotic velocity of the spacecraft, which lead to landing in a given region of the surface.

As it can be seen from the map in Fig. 10, there are two different trajectories leading to landing in Vellamo-South, so it became possible to reach this region at launch on the chosen date from the launch window in 2031. It is worth emphasizing here that if the condition (7.1) is violated on the chosen launch date, the only one landing circle may exist, or no landing circles may exist at all, i.e. the required 1:1 resonant trajectory may not exist.

### 2.2. Estimation of the total area attainable for landing

The total area on the Venus surface attainable for landing when using the proposed method is formed by all possible landing circles corresponding to various values of the $\gamma$ angle. Fig. 11 illustrates the formation of the total attainable landing area by landing circles corresponding to various feasible 1:1 resonant orbits.

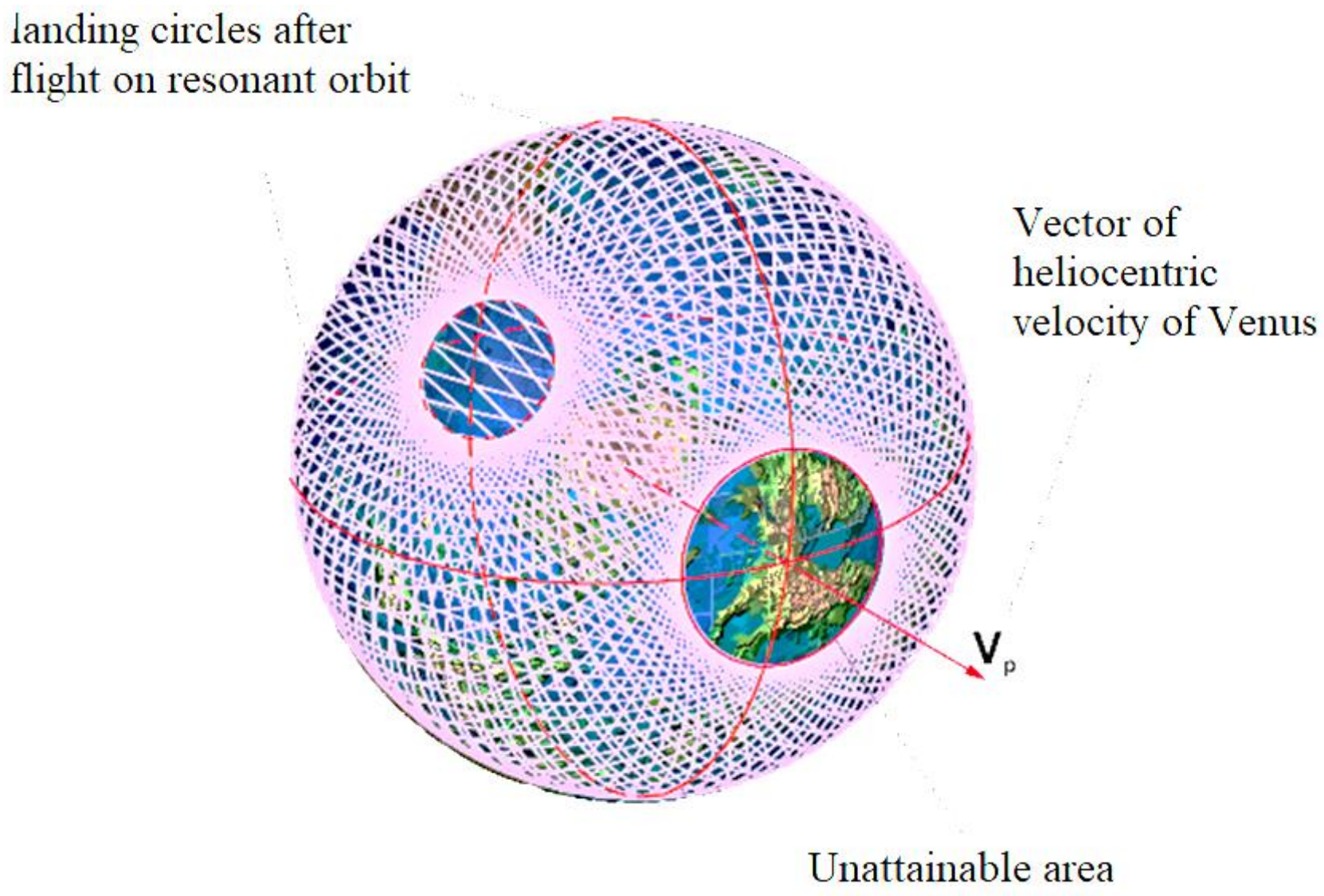

**Fig. 11.** Formation of the total attainable landing area by all possible landing circles at the angle of the entry to the atmosphere of Venus equal to 25 deg. Individual landing circles are shown in lilac.

Fig. 12 presents the results of the application of the proposed method to the mission design at launch windows in 2029-2037. Note, that the time interval considered as the launch window was taken as ±2.5 weeks from the optimal launch date (see Table 1) by the criterion (11). The Figure 12 shows the maps of Venus, on which the areas unattainable for landing are shadowed for the selected dates of launch from the Earth and landing on Venus. Note that due to orbital motion and own rotation of Venus unattainable areas are shifted for different launch-landing dates.

As a result of the analysis of unattainable areas on the surface of Venus within the launch in considered launch windows it is possible to highlight the area of the surface of Venus which remains unattainable for landing at any of the launch dates within the launch window. In Fig. 12 these totally unattainable areas are shaded in yellow. Usually there are two such regions which coordinates on the surface of Venus differ by 180 degrees in longitude, i.e. these areas are located in the vicinity of the point of intersection of the vector of the orbital velocity of Venus with its surface.

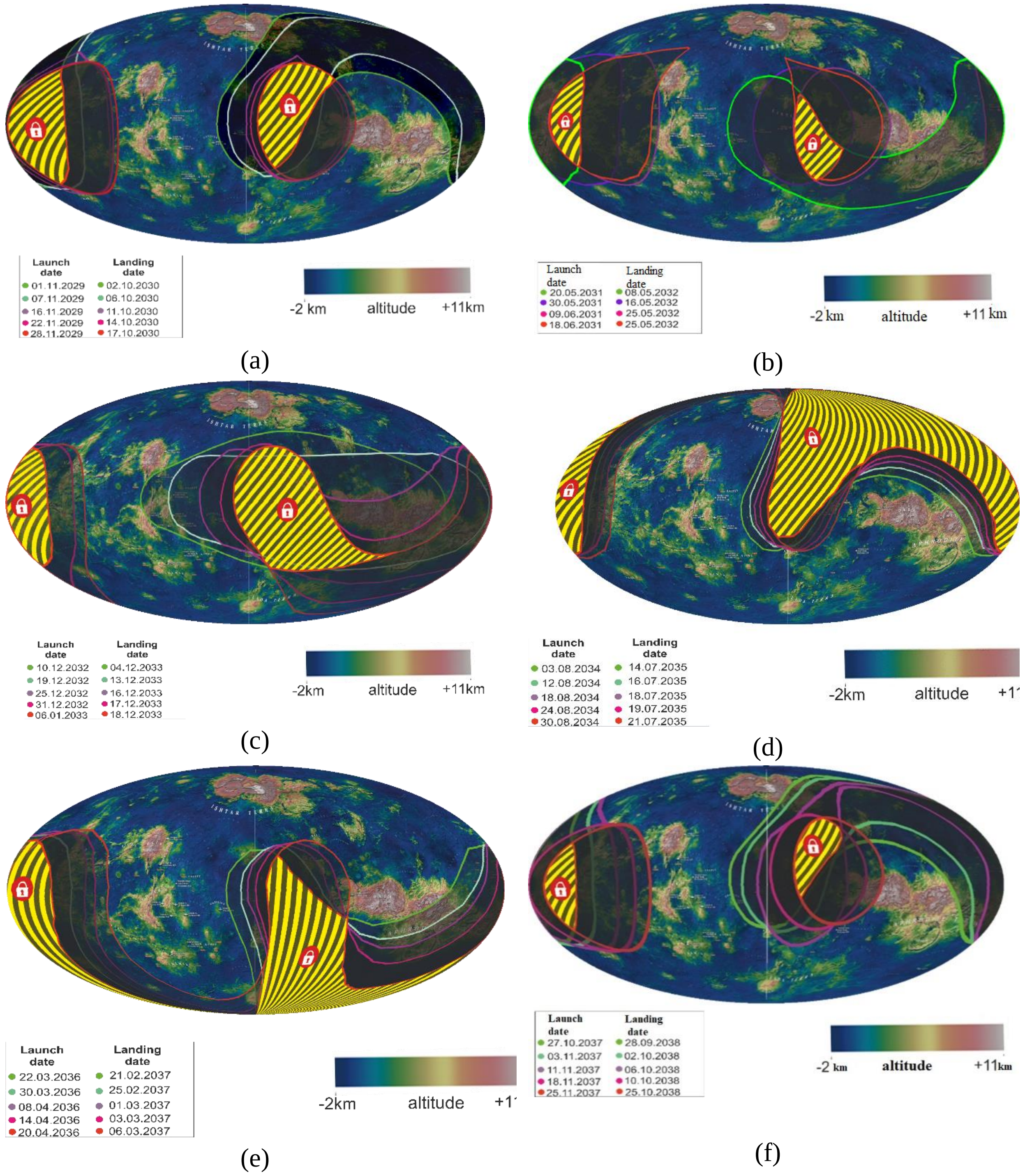

**Fig. 12.** Areas unattainable for landing for different landing dates (are shadowed on the Venus map) and their shifting on the Venus surface depending on the launch date in (a) 2029; (b) 2031; (c) 2032; (d) 2034; (e) 2036; (f) 2037. Totally unattainable areas are shaded in yellow. The areas are plotted on the Venus map in an equilateral Mollweide projection, the central meridian is aligned with zero for Venus (vicinity of the Alpha region). Dates on the figures are given in the format dd.mm.yyyy.

Existence and size of the areas totally unattainable for landing depend on the value of the entry angle into the Venusian atmosphere. In (Eismont et al., 2022b), a detailed analysis of the effect of the entry angle into the atmosphere on the arising areas totally unattainable for landing is made. In the mentioned work, it was shown that if an atmospheric entry angle is of 35 degrees, the radius of the landing circle is close to 90 degrees, and, therefore, it is possible to avoid the appearance of totally unattainable areas (if the condition (7.1) is fulfilled). Because of different reasons, in real missions the selected entry angle is often less than 35 degrees, the radius of the landing circle turns out to be less than 90 degrees, and unattainable areas appear. Centers of such areas are the intersection points of the orbital velocity vector of Venus with its surface (a simple example of this is shown at Fig. 11).

It is possible to avoid the appearance of totally unattainable areas since Venus moves in its orbit shifting by about 1.5 degrees per day. Using this natural property, it is possible to increase the flight time of the spacecraft, or rather the flight time before the gravity assist maneuver, thus the unattainable area shifts, and the areas previously unattainable become attainable (Eismont et al. 2022b).

### 2.3. Refining the spacecraft trajectory parameters in the design of trajectory to Venus with landing in the Vellamo-South region

Above the trajectory to Venus with landing in Vellamo-South region using the proposed method was obtained. In this section we compare that trajectory with the one refined in a more complicated model of motion, considering gravity influences of other celestial bodies as well as the solar pressure. The refined trajectory was calculated using the GMAT NASA Software[8].

Let us present the calculated parameters of the Earth-Venus-Venus trajectory of the spacecraft (see Table 3) obtained in the framework of the patched conic approximation and then refined.

**Table 3.** Comparison of the parameters of the Earth-Venus-Venus spacecraft trajectory obtained in the framework of the patched conic approximation and the refined trajectory.

| | Patched conic approximation | Complicated model |
|---|---|---|
| Launch date | Jun 03, 2031 | Jun 04, 2031 |
| $\Delta V_0$, km/s | 3.81 | 3.83 |
| $r_{\pi 0}$, (km), $i_0$ (deg.) | 6571, 51.6 | 6571, 51.6 |
| Date of the Venus flyby | 08.10.2031 | 08.10.2031 |
| BdotR, BdotT (km) | -4113.15; -33248.99 | |
| The spacecraft state vector at the periapsis of the hyperbolic flyby trajectory | $r_\pi = \begin{bmatrix} -12091.8 \\ -1685.0 \\ 3210.9 \end{bmatrix}; V_\pi = \begin{bmatrix} -0.9363 \\ 7.6718 \\ 0.4997 \end{bmatrix}$ | $r_\pi = \begin{bmatrix} -12030.3 \\ -1546.6 \\ 3348.4 \end{bmatrix}; V_\pi = \begin{bmatrix} -0.8394 \\ 7.6854 \\ 0.5797 \end{bmatrix}$ |
| $\Delta V_\pi$, km/s | 0 | $1.2\times10^{-2}$ |
| $r_\pi$, km | 12624 | 12585 |
| $V_r = V_r^- = V_r^+$, km/s | 2.91 | 2.92 |
| $RA^-$, $DEC^-$ deg. | 145.3; 13.5 | 144.7; 14.4 |
| $RA^+$, $DEC^+$ deg. | 49.6; -8.5 | 49.1; -8.6 |
| Date of landing to the Venus surface | May 20, 2032 | May 19, 2032 |
| Angular shift relative to the landing point Vellamo-South, deg | - | 0.27 |
| Linear shift relative to the landing point Vellamo-South, km | - | 27.5 |

Note. BdotR, BdotT are the parameters of the B-plane calculated as in (Vallado 2016). $\Delta V_\pi$ – impulse applied to the spacecraft at Venus flyby. $RA^-$, $DEC^-$ are right ascension and declination of the vector of the incoming asymptotic velocity of the spacecraft in the hesperocentric ecliptic coordinate system; $RA^+$, $DEC^+$ are right ascension and declination of the vector of the outgoing asymptotic velocity of the spacecraft in the hesperocentric ecliptic coordinate system.

Fig. 13 shows landing circles constructed when calculating the flight trajectory using the patched conic approximation and the complicated model. Let us also introduce the parameter $\beta$ (see Fig.

[8] URL https://software.nasa.gov/software/GSC-17177-1 (access date July 07, 2022)

13) which is the angle between directions to the very south point of the landing circle (-65°S, 122°E) and to the current landing point counted counterclockwise.

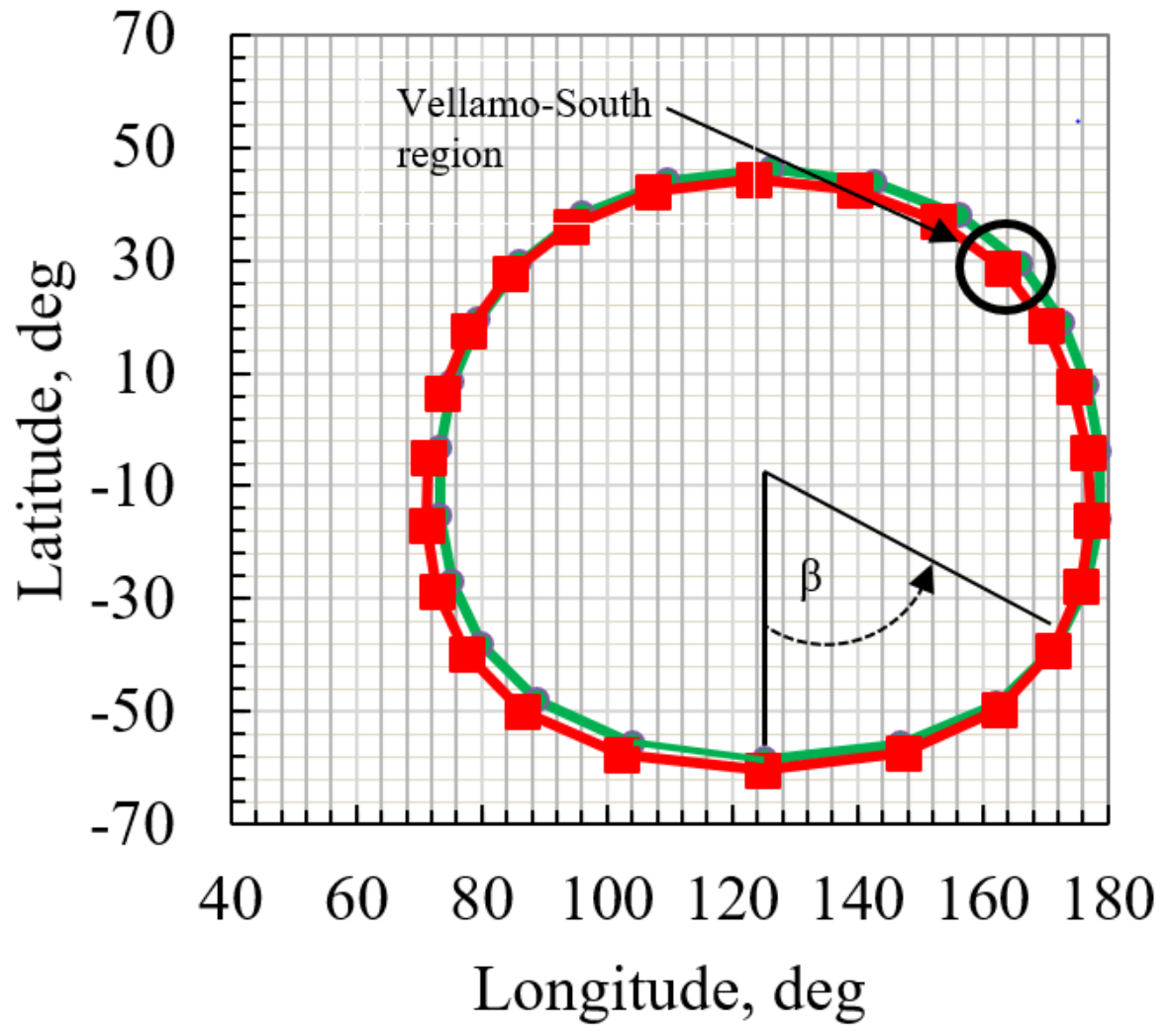

**Fig. 13.** Landing circles constructed in the patched conic approximation (yellow-green dots) and in the complicated model (red dots), respectively

Let us show the landing date shift as well as angular Δν and longitudinal Δ*L* shifts obtained in the result of the described refining of the trajectory parameters vs the *β* parameter (see Fig. 14 (a) and (b)).

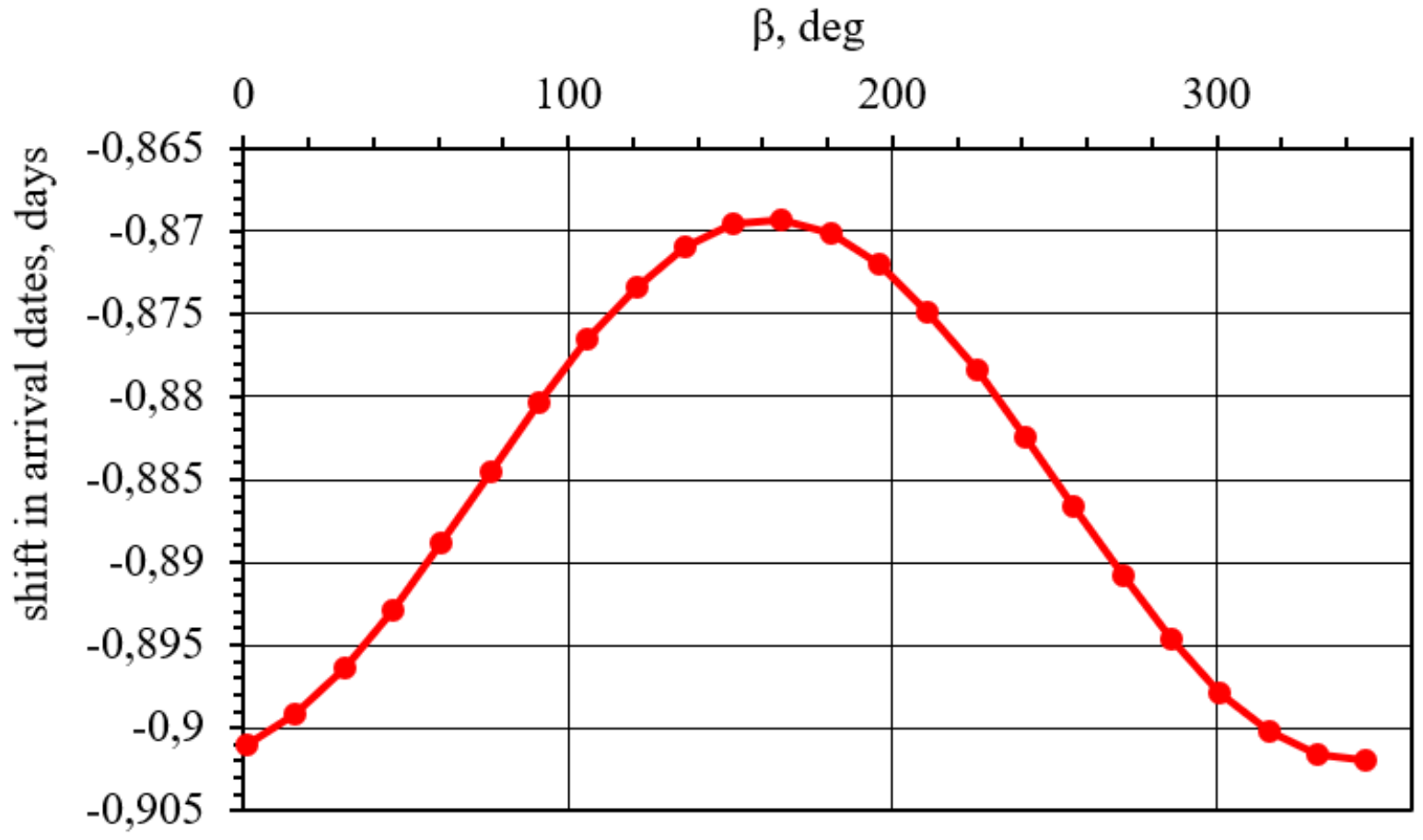

(a)

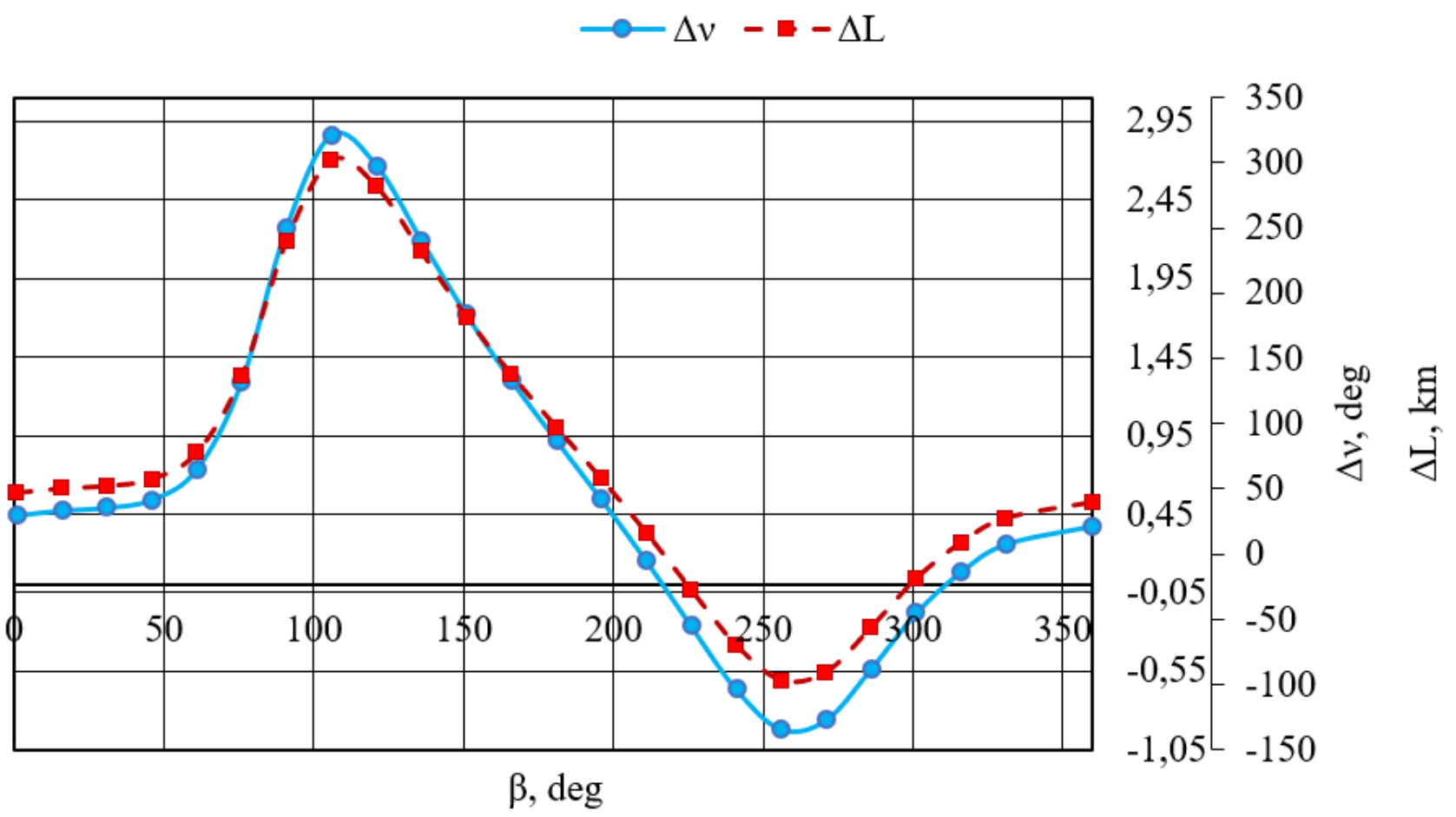

(b)

**Fig. 14.** Shift of the landing date (a) and landing points (b) after refining the trajectory parameters using the complicated model.

Fig. 14 (a) shows that the landing date shifts by about 1 day towards earlier landing dates and varies from 0.87 to 0.9 days. Fig. 14 (b) shows the linear displacement of the landing point. Depending on the selected landing point, it can be seen that the maximum displacement of the landing point from the nominally defined one is about 300 km.

From Fig. 14 it can be seen that the refinement of the parameters of the spacecraft trajectory leads only to a slight shift in the landing circle, which is explained by the approach Venus 1 day earlier than was obtained in the patched conics model.

## Summary and conclusions

In this paper, we proposed a method to design a trajectory of a flight to Venus, ensuring landing in a given region on its surface. The effectiveness of this method is shown in comparison with the construction of the direct flight trajectory from the Earth to Venus.

The effectiveness of the proposed method is due to the peculiarities of orbital motion and own rotation of Venus, as well as limited capabilities of the spacecraft launch from the Earth, which is possible only every 19 months. As a result of this, during a direct flight, a part of the planet's surface which is attainable for landing does not exceed 5% of the entire surface. The use of the proposed method makes it possible to land almost anywhere on the entire surface, with the exception, perhaps, small areas. In the problem of designing a mission to other planets (Mercury, Mars, giant planets), the effectiveness of the method may significantly decrease. As for the planets of the terrestrial group, due to the less dense atmosphere of Mercury and the faster (compared to Venus) rotation of Mars, designing a trajectory to these planets using the proposed method is less effective than using traditionally established methods. For giant planets, it should be noted that their orbital period significantly exceeds the Venusian one, and therefore the use of resonant orbits turns out to be ineffective. However, in the design of a flight trajectory in the systems of moons of giant planets, the application of the proposed method can show high efficiency, since it does not depend on the nature of the moons' own rotation, which is synchronous for most of the large moons. Thus, the application of the proposed method in the design of missions to the giant planet moons, providing landing on their surface, is a promising direction for further research.

The dependence of the $\mathbf{V}_r^+$ vector on the latitude and longitude of the landing point given in the current paper makes it easy to establish a connection between the required landing point on the surface of Venus and the resonant orbit leading to this point, and also to check whether this point of the surface is in principle attainable for landing.

The paper provides an example of the application of the method for constructing a flight trajectory with landing in the scientifically important Vellamo-South region. It is shown that landing in this area as part of a direct flight at launch in 2031 is impossible. However, the application of the proposed method allows us to construct two different trajectories leading to landing in this area. The selection of a specific trajectory from these two possible ones depends on additional restrictions that arise, among other factors, due to considering the peculiarities of motion of an orbiter (if the mission provides for its presence).

The analysis of the Venusian surface regions that are not attainable for landing at the mission launch from 2029 to 2037 shows that in all cases the proposed method allows landing in almost any area on the surface, and the superposition of unattainable areas within one launch window and, moreover, within the entire considered period of launch dates shows that any point on the surface of Venus may be made attainable for landing.

A comparison of the approximate flight trajectory constructed within the framework of the proposed method with the trajectory refined using the complicated model of the spacecraft motion shows the effectiveness of the proposed method, since the difference in the costs of the characteristic velocity $\Delta V_0$ does not exceed 2% compared to the patched conic model, however, the launch date shifts by one day, the offset of the landing time on Venus is on average also one day. At the same time, the deviation of the landing point in the Vellamo-South region when integrating the refined trajectory is 27.5 km relative to the one obtained in the method of patched conics. All this suggests that the results obtained in the simplified model are a good initial approximation to the construction of the flight trajectory to Venus, sufficient for carrying out design estimates.

## Appendix A

This appendix provides evidences for some of the statements made in section 1.2 of this paper.

**Lemma 1.** The shortest turn of the vector $\mathbf{V}_r^{-LO}$ to the position $\mathbf{V}_r^{+LO}$ belonging to the cone of asymptotic velocities that ensure the transfer of the spacecraft to the resonant orbit is a turn in the plane formed by the vectors $\mathbf{V}_p^{LO}$ and $\mathbf{V}_r^{-LO}$.

***Proof.*** Consider the vectors $\mathbf{V}_p^{LO}$, $\mathbf{V}_r^{-LO}$ and $\mathbf{V}_r^{+LO}$ in the local orbital system $\xi\eta\zeta$ (see Fig. 6). In this coordinate system projections of the vectors $\mathbf{V}_p^{LO}$ and $\mathbf{V}_r^{-LO}$ are defined as follows:

$$\mathbf{V}_r^{-LO} = V_r \left\{\cos\delta,\ \sin\gamma_0 \sin\delta,\ \cos\gamma_0 \sin\delta\right\}^T, \tag{a}$$

$$\mathbf{V}_p^{LO} = V_p \left\{1,\ 0,\ 0\right\}^T,$$

where $\gamma_0 = \arccos\dfrac{\left(\mathbf{V}_r^{-LO}\right)_{\eta\zeta}\cdot\mathbf{n}}{\left|\left(\mathbf{V}_r^{-LO}\right)_{\eta\zeta}\right|}$, $\mathbf{n} = \{0,0,1\}^T$; $\delta = \arccos\dfrac{\mathbf{V}_r^{-LO}\cdot\mathbf{V}_p^{LO}}{|\mathbf{V}_r^{-LO}||\mathbf{V}_p^{LO}|}$.

Obviously, the projection $\left(\mathbf{V}_r^{-LO}\right)_{\eta\zeta}$ lies on the intersection line of the planes formed by the vectors $\mathbf{V}_p^{LO}$ and $\mathbf{V}_r^{-LO}$, therefore, then we set the position of the plane by an $\gamma_0$ angle relative to the $\zeta$ axis.

The vector of the asymptotic velocity after the gravity assist maneuver can be written as:

$$\mathbf{V}_r^{+LO} = V_r \left\{\cos\Phi,\ \sin\gamma \sin\Phi,\ \cos\gamma \sin\Phi\right\}^T, \tag{b}$$

where $\Phi$ is an angle between $\xi$-axis and $\mathbf{V}_r^{+LO}$.

The angle of turn of the asymptotic velocity vector of the spacecraft from the $\mathbf{V}_r^{-LO}$ position to the $\mathbf{V}_r^{+LO}$ position is determined as follows:

$$\alpha_1 = \arccos\frac{\mathbf{V}_r^{-LO}\cdot\mathbf{V}_r^{+LO}}{|\mathbf{V}_r^{-LO}||\mathbf{V}_r^{+LO}|}. \tag{c}$$

Substituting (a) and (b) into (c), we get:

$$\alpha_1 = \arccos\left(\cos\Phi\cos\delta + \sin\Phi\sin\delta\sin\gamma_0\sin\gamma + \sin\Phi\sin\delta\cos\gamma_0\cos\gamma\right). \tag{d}$$

Since the transition to the orbit of the required resonance is considered, the angles $\Phi$ and $\delta$ are constant, since the first of them determines the required resonance line, and the second is determined by the orientation of the vector of the incoming asymptotic velocity and does not change during the gravity assist maneuver.

Then to find the shortest turn angle, take the derivative of $\alpha$ by $\gamma$:

$$\frac{d\alpha}{d\gamma} = -\frac{B}{\sqrt{1-C^2}}\left(\cos\gamma\sin\gamma_0 - \sin\gamma\cos\gamma_0\right),$$

where

$$C = A + B\left(\sin\gamma_0 \sin\gamma + \cos\gamma_0 \cos\gamma\right),$$

$$A = \cos\Phi\cos\delta$$
$$B = \sin\Phi\sin\delta \; .$$

Extremes of the function (6) may be found from the condition $\left.\frac{d\alpha}{d\gamma}\right|_{\gamma=\gamma_{extr}} = 0$ which can be written as

$$-\frac{B}{\sqrt{1-C^2}}\left(\cos\gamma\sin\gamma_0 - \sin\gamma\cos\gamma_0\right) = 0 \, .$$

Since $-\frac{B}{\sqrt{1-C^2}} \neq 0$, therefore $\left(\cos\gamma\sin\gamma_0 - \sin\gamma\cos\gamma_0\right) = 0$, or $\sin\gamma_0 - \cos\gamma_0 \mathrm{tg}\gamma = 0$, or $\mathrm{tg}\gamma = \mathrm{tg}\gamma_0 \Rightarrow \gamma_{extr} = \gamma_0 + \pi n, \;\; n = 1, 2$, since $\gamma \in \left[0, 2\pi\right]$.

Thus, two solutions for the $\gamma$ angle are obtained (since $\gamma_{extr} = \gamma_0 + 2\pi = \gamma_0$):

$$\gamma_{extr} = \gamma_0 \text{ and } \gamma_{extr} = \gamma_0 + \pi \, . \tag{e}$$

Since in both cases the extremum coincides with the line of intersection between the plane formed by **V**p and $\mathbf{V}_r^-$ and the plane normal to **V**p, therefore, the lemma is proved and the shortest angle of turn really lies in the specified plane.

*Remark 1.* Two solutions of $\gamma$ define two turn angles, one of which determines the smallest turn angle, and the second one determines the largest one. The corresponding angles can be obtained by substituting (d) in (e):

$$\begin{cases} \alpha_{\min} = \Phi - \delta \\ \alpha_{\max} = \Phi + \delta \end{cases} . \tag{f}$$

*Remark 2.* If *δ=0* then $\alpha_{\min} = \alpha_{\max} = \Phi$ .

**Theorem 1.** For the existence of an impulse-free transition of the spacecraft to a resonant orbit using a gravity assist maneuver near a planet, at a given spacecraft asymptotic velocity $V_r = |\mathbf{V}^-{}_r| = |\mathbf{V}^+{}_r|$, it is necessary and sufficient to fulfill conditions (4) and (5).

***Proof.***

The necessity follows from the remark to Lemma 1, in which the expression for the shortest angle of turn $\alpha_{\min} = \Phi - \delta$ was obtained. Since $\Phi$ is the angle between $\mathbf{V}_p^{LO}$ and $\mathbf{V}_r^{+LO}$, therefore, it can be defined as follows:

$$\Phi = \arccos\frac{\left(V_a^+\right)^2 - V_r^2 - V_p^2}{2V_rV_p} ,$$

where $V_a^+$ is the spacecraft velocity on a heliocentric orbit after the gravity assist maneuver.

Since the arccosine argument can vary from -1 to 1, we can write $-1 \le \frac{\left(V_a^+\right)^2 - V_r^2 - V_p^2}{2V_rV_p} \le 1$, or $V_p^2 - 2V_rV_p + V_r^2 \le \left(V_a^+\right)^2 \le V_p^2 + 2V_rV_p + V_r^2$, or, finally,

$$|V_p - V_r| \le V_a^+ \le V_p + V_r \, .$$

We obtained the inequality (5). Thus, in order to be able to rotate the vector $\mathbf{V}_r^-$ to the $\mathbf{V}_r^+$ position, it is necessary that the above relation is fulfilled.

Sufficiency is proved quite simply, relying on the main conclusion from Lemma 1. Since the necessary condition (5) is fulfilled, $\exists \Phi \Rightarrow \exists \alpha_{\min} = \Phi - \delta$. As shown in Lemma 1, this angle is the smallest among the required turn angles of the asymptotic velocity vector from $\mathbf{V}_r^-$ to $\mathbf{V}_r^+$. If the smallest angle of turn is less than the maximum natural angle of turn, and at the same time $|\mathbf{V}_r^-| = |\mathbf{V}_r^+|$, it is obvious that there is at least one impulse-free transition of the spacecraft to a resonant orbit by means of a gravity assist maneuver.